\documentclass[a4paper,fleqn,usenatbib]{mnras}

\usepackage{mathptmx}
\usepackage{txfonts}

\usepackage[T1]{fontenc}
\usepackage{ae,aecompl}
\usepackage{multicol}
\usepackage{graphicx}
\usepackage{color}
\usepackage{graphicx}
\usepackage{amssymb}
\usepackage{rotating}
\usepackage{longtable}
\usepackage{lscape}
\usepackage{multirow}
\usepackage{multicol}

\newcommand{\figref}{Figure~\ref}
\newcommand{\tabref}{Table~\ref}

\newcommand{\secref}{Section~\ref}

\title[The globular clusters NGC\,1851, NGC\,1904, NGC\,2298 and NGC\,2808]{Tails and streams around the Galactic globular clusters NGC\,1851, NGC\,1904, NGC\,2298 and NGC\,2808}

\author[J. A. Carballo-Bello et al.]{Julio A. Carballo-Bello$^{1,2}$\thanks{E-mail: jcarballo@astro.puc.cl}, David Mart\'inez-Delgado$^{3}$, Camila Navarrete$^{1,2}$,\newauthor 
M\'arcio Catelan$^{1,2}$, Ricardo R. Mu\~noz$^{4}$, Teresa Antoja$^{5}$ \& Antonio Sollima$^{6}$\\
$^{1}$Instituto de Astrof\'isica, Facultad de F\'isica, Pontificia Universidad Cat\'olica de Chile, Av. Vicu\~na Mackenna 4860, 782-0436 Macul,\\ 
Santiago, Chile\\
$^{2}$Millenium Institute of Astrophysics, Santiago, Chile\\
$^{3}$Astronomisches Rechen-Institut, Zentrum f\"ur Astronomie der Universit\"at Heidelberg, M\"onchhofstr. 12-14, D-69120 Heidelberg, Germany\\
$^{4}$Departamento de Astronom\'ia, Universidad de Chile, Camino El Observatorio 1515, Las Condes, Santiago, Chile\\
$^{5}$Dept. FQA, Institut de Ciencies del Cosmos (ICCUB), Universitat de Barcelona (IEEC-UB), Marti Franques 1, E08028 Barcelona, Spain\\
$^{6}$INAF Osservatorio Astronomico di Bologna, via Ranzani 1, I-40127 Bologna, Italy\\
}

\date{Accepted XXX. Received YYY; in original form ZZZ}

\pubyear{2016}

\begin{document}
\label{firstpage}
\pagerange{\pageref{firstpage}--\pageref{lastpage}}
\maketitle

\begin{abstract}
We present DECam imaging for the peculiar Galactic globular clusters NGC\,1851, NGC\,1904 (M\,79), NGC\,2298 and NGC\,2808. Our deep photometry reveals that all the clusters have an important contribution of stars beyond their King tidal radii and present  tails with different morphologies. We have also explored the surroundings of the clusters where the presence of the Canis Major overdensity and/or the low Galactic latitude Monoceros ring at $d_{\odot} \sim 8$\,kpc is evident. A second stellar system is found at $d_{\odot} \sim 17$\,kpc and spans at least $18\,{\rm deg} \times 15\,{\rm deg}$ in the sky. As one of the possible scenarios to explain that feature, we propose that the unveiled system is part of Monoceros explained as a density wave moving towards the outer Milky Way. Alternatively, the unveiled system might be connected with other known halo substructures or associated with the progenitor dwarf galaxy of NGC\,1851 and NGC\,1904, which are widely considered accreted globular clusters.
\end{abstract}


\section{Introduction}

Globular clusters (GCs) are excellent tracers of the hierarchical formation of galaxies. Bright GCs are observed at large distances and represent a powerful tool to understand how the stellar halos in the Local Universe are assembled and test the predictions made by the $\Lambda$-Cold Dark Matter simulations \cite[e.g.][]{Font2011}. For instance, an important number of GCs in M\,31 seem to be aligned with the progeny of tidal streams unveiled in recent years around that galaxy \citep{Mackey2010,Mackey2013,Huxor2014,Veljanoski2014}, showing the complex formation history of its outer regions. The Galactic GC system also reflects the dual formation process of the Milky Way with a component of clusters likely accreted or captured from other dwarf stellar systems already assimilated by our Galaxy \citep[e.g.][]{Zinn1993, Marin-Franch2009,Leaman2013,Zaritsky2016}. Indeed, between 9 and 20 Galactic GCs have been associated with the Sagittarius tidal stream \citep{Majewski2003,Belokurov2006,Koposov2012} based on their positions with respect to the observed or predicted path of that halo substructure \citep[e.g.][]{Bellazzini2002,Martinez-Delgado2002,Bellazzini2003,Law2010b,Carballo-Bello2014,Sbordone2015}. Recently, the detection of the old leading arm of that stream around Whiting\,1 has been put forward by \cite{Carballo-Bello2017}.

The Monoceros ring, a vast stellar structure at low Galactic latitudes \citep{Newberg2002,Yanny2003,Juric2008,Conn2012,Slater2014,Morganson2016}, might also contain many GCs, regardless of the origin of those clusters and that of the ring. Several scenarios have been proposed to explain the formation of such a remarkable structure: accretion of a dwarf galaxy \citep[e.g.][]{Conn2005,Penarrubia2005,Sollima2011,Carballo-Bello2015}, propagation of a density wave through the Milky Way disc \citep[e.g.][]{Kazantzidis2008,Purcell2011,Xu2015}, the internal structure of the Galaxy (warp, flare or spiral arm)  along that line of sight \citep{Momany2006,Hammersley2011,Lopez-Corredoira2014}, and the existence of a dark matter caustic ring \citep{Natarajan2007}. One of the main difficulties to fully understand the nature of the Monoceros ring is the lack of a clear progenitor galaxy as in the case of the Sagittarius tidal stream. The  Canis Major overdensity \citep{Martin2004} has been proposed as the hypothetical progenitor of the Monoceros ring although evidence has been presented both in favor and against its extra-Galactic nature  \citep{Momany2006,Martinez-Delgado2005,Moitinho2006,Butler2007,Mateu2009}.  

Among the Galactic GCs likely inmersed in Monoceros, NGC\,1851 and NGC\,1904 (M\,79) are two of the main candidates to be accreted clusters. Together with NGC\,2298 and NGC\,2808, those globulars are confined in a sphere of radius $R \sim 6$\,kpc that resembles the spatial distribution of the clusters M\,54 (NGC\,6715), Terzan\,7, Terzan\,8 and Arp\,2, a group of GCs located in the main body of the Sagittarius dwarf galaxy \citep{Bellazzini2004}. Moreover, \cite{Forbes2010} found that these globulars fit in the branch of accreted clusters in the age-metallicity relation derived for the Milky Way GC system. Interestingly, as noted by \cite{Martin2004}, NGC\,1851, NGC\,1904, NGC\,2298 and NGC\,2808 are found close to the region populated by the controversial Canis Major overdensity. \cite{Penarrubia2005} considered that possibility and compared the position of those GCs with that of their predicted orbit for the ring. Despite the spatial coincidence, NGC\,1851, NGC\,1904 and NGC\,2298 seem to move on highly eccentric orbits (retrograde in the case of NGC\,2298), which are not compatible with a low Galactic latitude feature as Monoceros. 
  
Independently of the hypothetical association of this group of GCs with Monoceros, they may be considered peculiar members of the Galactic GC population. NGC\,1851 hosts at least two well-differentiated stellar populations, which manifest in its colour-magnitude diagram (CMD) \citep{Milone2008,Milone2009,Zoccali2009,Piotto2012,Cummings2014}. Evidence of the presence of multiple populations has been also reported based on the chemical abundances derived for cluster members \citep[e.g.][]{Carretta2011,Gratton2012}. The abundance pattern of NGC\,1851 has been considered evidences of its extra-Galactic origin and peculiar birth conditions \citep{Carretta2011,Bekki2012}. \cite{Olszewski2009} argued that the main sequence (MS) population of NGC\,1851 might be present at distances up to 75\,arcmin (250\,pc) from the cluster centre. Spectroscopy has confirmed that a large number of stars with velocities similar to that of NGC\,1851 are found beyond its tidal radius \citep{Kunder2014,Marino2014,Navin2015}. In addition, \cite{Sollima2012} derived radial velocities for a sample of stars selected from the \cite{Carballo-Bello2014} catalogs and found an unexpected component in the radial velocity distribution, which is associated neither with the cluster nor with the Milky Way populations. The latter result may indicate that NGC\,1851 is surrounded by a stellar system in which the cluster might have been formed.  

Chemical abundances provide hints about the tentative presence of multiple populations in NGC\,1904 \citep[e.g.][]{Orazi2015}, but no additional evolutionary branches in its CMD have been reported. NGC\,1904 hosts extended horizontal branch stars which might be considered one of the characteristics of captured GCs \citep{Lee2007}. NGC\,2808 hosts at least 5 stellar populations \citep{Piotto2007,Milone2015} which are also revealed through detailed chemical abundances analysis \citep[e.g.][]{Dantona2005,Marino2014}. On the other hand, the poorly studied cluster NGC\,2298 is the only member of this group of GCs for which the existence of multiple stellar populations has not yet been reported in the literature, although a tentative secondary population is observed in the CMD derived by \cite{Piotto2015}. 

NGC\,1851, NGC\,1904, NGC\,2298 and NGC\,2808 are located in a region of the Milky Way with substructures possibly originated by the accretion of an already disrupted system. Moreover, the detection of a hypothetical cold stream along the line of sight of NGC\,1851 by \cite{Sollima2012} and the surprising extended stellar halo around that cluster \citep{Olszewski2009,Kuzma2017} encourage us to explore the area between and around NGC\,1851 and NGC\,1904. The structure of these GCs might also provide clues about the overall mass distribution of the Milky Way. Indeed, the shape and  distribution of tidal tails emerging from Galactic GCs may contain valuable information about the potential and dark matter distribution of the Galaxy \citep[e.g.][]{Law2010a}. Moreover, it has been suggested that the gaps observed in these stellar structures may be produced by the presence of small dark matter halos in the Milky Way \citep[.e.g.][]{Erkal2015}.  

In this paper, we have used deep  photometry to study the structure of these GCs and search for underyling stellar systems around them. 

\section{Observations and methodology}

\begin{table}
\small
\begin{centering}
\begin{tabular}{lllll}
Field $\#$ & \emph{l}($^{o}$) & \emph{b}($^{o}$) & $g$ t$_{\rm exp}$ (s) & $r$ t$_{\rm exp}$ (s)\\
\hline
\\ 
1 (NGC\,1851) & 243.62 & -34.83 & 2$\times$300 & 3$\times$200 \\ 
2 & 241.02 & -33.97 & 2$\times$300 & 3$\times$200 \\
3 & 238.41 & -33.12 & 2$\times$300 & 3$\times$200 \\
4 & 235.81 & -32.26 & 2$\times$300 & 3$\times$200 \\
5 & 233.21 & -31.41 & 2$\times$300 & 3$\times$200 \\
6 & 230.61 & -30.55 & 2$\times$300 & 3$\times$200 \\
7 (NGC\,1904) & 228.00 & -29.69 & 2$\times$300 & 3$\times$200 \\
8 & 200.00  & -32.19 & 2$\times$300 & 3$\times$200 \\
9 &  235.87 & -26.00 & 2$\times$300 & 3$\times$200 \\
10 &  235.87 & -18.00 & 2$\times$300 & 3$\times$200 \\
11 (NGC\,2298) &  245.63 & -16.00 & 300 & 200 \\
12 (NGC\,2808) &  282.19 & -11.25 & 300 & 200 \\
\\
\hline
\end{tabular}
\caption{Position of the fields observed in Galactic coordinates and the exposure times for the bands $g$ and $r$. Fields 1, 7, 11 and 12 contain, respectively, the GCs NGC\,1851, NGC\,1904, NGC\,2298 and NGC\,2808.}
\label{Observationstable}
\end{centering}
\end{table}

  \begin{figure}
     \begin{center}
      \includegraphics[scale=0.3]{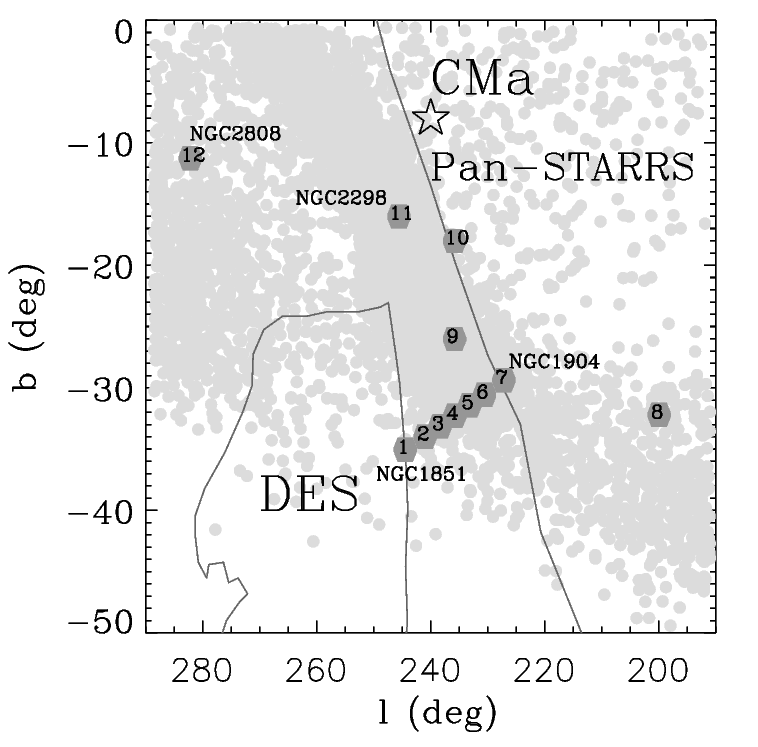}
      \caption[]{Projected position of the fields observed with DECam (indicated by the numbers). The Monoceros ring path, as predicted by the \cite{Penarrubia2005} model, has been included as a reference for this work. Grey circles correspond to particles from the simulation. The footprint of Pan-STARRS and DES are also indicated as solid lines while the star corresponds to the central coordinates of the main body of the Canis Major overdensity.} 
\label{monoceros}
     \end{center}
   \end{figure}

In this work we have used the Dark Energy Camera (DECam), which is mounted at the prime focus of the 4-m Blanco telescope at Cerro Tololo Inter-American Observatory (CTIO). DECam provides a 3\,deg$^{2}$ field of view with its 62 identical chips with a scale of 0.263\,arcsec\,pixel$^{-1}$ \citep{Flaugher2015}. The observations were executed on January 08, 2014 under the proposal ID 2013B-0615. The observing run consisted of one single dark night, with clear sky conditions and all the exposures were initiated when their airmasses were smaller than 1.2. In order to cover the area around and between the GCs NGC\,1851 and NGC\,1904, we observed 7 equidistant DECam fields along the line connecting both clusters. We also included 3 fields to explore the surroundings of the NGC\,1851/1904 system: one of them at similar Galactic latitude but at $\ell$ = 200$^{\circ}$ and other two fields with the same Galactic latitude as the central field between the globulars but with $b$ = -26$^{\circ}$ and -18$^{\circ}$. The exposure times were $2\times300$ and $3\times200$\,s for the $g$ and $r$ bands, respectively. Additional fields were included to observe NGC\,2298 and NGC\,2808, with exposure times 300 and 200\,s for the $g$ and $r$ bands (see \tabref{Observationstable}). Seeing conditions were stable during the night, so no important distortions between fields are found. As shown in \figref{monoceros}, field\,1 is found within the footprint of the Dark Enery Survey \citep[DES;][]{DES2016}, while fields 7, 8 and 10 are located in the area of the sky covered by the Panoramic Survey Telescope and Rapid Response System\,1 \citep[Pan-STARRS\,1;][]{Chambers2016} $3\pi$ survey.

The images were processed by the DECam Community Pipeline \citep{Valdes2014} and accesed via the NOAO Science Archive. Photometry was obtained from the summed images with the PSF-fitting algorithm of \textsc{DAOPHOT II/ALLSTAR}
\citep{Stetson1987}. The final catalog only includes stellar-shaped objects with |sharpness| < 0.5. We observed 4 Sloan Digital Sky Survey (SDSS) fields at different airmasses to derive the atmospheric extinction coefficients and transformation between the instrumental magnitudes and the $ugriz$ system.  In the following analysis all magnitudes have been dereddened using the $E(B-V)$ maps by \cite{Schlafly2011}.

\begin{table*}
\centering
\begin{tabular}{lcccrcrcccccccc}

\multicolumn{9}{l}{} & \multicolumn{2}{c}{King} &\multicolumn{2}{c}{Power Law} & \multicolumn{2}{c}{Wilson} \\
Cluster & R.A. & Dec & $\ell$ & $b$ & $d_{\odot}$ & [Fe/H] & $age$ & v$_{\rm r}$ & $c$ & $r_{\rm c}$ & $r_{\rm 0}$ & $\gamma$ & $c$ & r$_{\rm c}$ \\
       & [hh:mm:ss] & [dd:mm:ss]  &  [deg] & [deg] & [kpc] & & [Gyr] & [km\,s\,$^{-1}$] &  & ['] & ['] &  &  & ['] \\
\hline
\\
NGC\,1851 & 05:14:06.7 & -40:02:47.6 & 244.51 & -35.03 & 12.1 & -1.18 & 10.0  & 320.5 & 2.11 & 0.07 & 0.19 & 3.0  & 2.15 & 0.10\\ 

NGC\,1904 & 05:24:11.1 & -24:31:29.0 & 227.23 & -29.35 & 12.9 & -1.60 & 11.0 & 205.8 & 1.75 & 0.16 & 0.19 & 2.7  & 1.86 & 0.19 \\ 

NGC\,2298 & 06:48:59.4 & -36:00:19.1 & 245.63 & -16.00 & 10.8 & -1.92 & 12.7 & 148.9 & 1.30 & 0.27 & 0.33 & 2.3  & 2.25 & 0.29\\ 

NGC\,2808 & 09:12:03.1 & -64:51:48.6 & 282.19 & -11.25 & 9.6 & -1.14 & 10.8 & 101.8 & 1.79 & 0.22 & 0.59 & 3.5  & 1.79 & 0.27 \\ 
\\									     
\hline
	  \end{tabular}

\caption[]{Coordinates, distances, metallicities and radial velocities for the GCs NGC\,1851, NGC\,1904, NGC\,2298 and NGC\,2808 \citep{Harris2010}. Absolute ages were taken from \cite{Forbes2010} while structural parameters are those derived in this work.}
\label{table1}

\end{table*}

\subsection{Completeness}

\begin{figure}
     \begin{center}
      \includegraphics[scale=0.35]{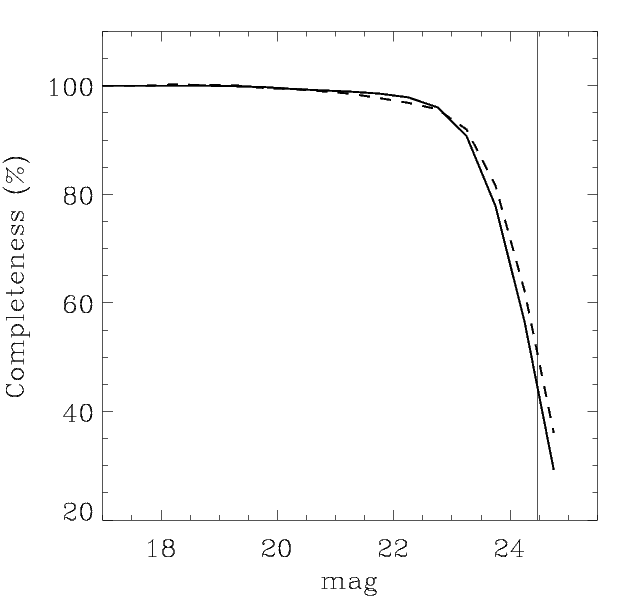}
      \caption[]{Completenes as a function of the magnitude for the field 3. The solid and dashed lines correspond to the results for the $g$ and $r$ bands, respectively. The vertical line indicates the magnitude where completeness drops to 50\% and indicates the limiting magnitude of our photometry.}
\label{completitud}
     \end{center}
   \end{figure}

In order to estimate the completeness of our photometric
catalogs, we have considered 5 central chips in fields 3, 4 and 5. Using \textsc{DAOPHOT\,II}, we have included in the images synthetic stars with magnitudes in the range $17 < g,r < 25$ and $0 < g-r < 1.5$, randomly distributed throughout the chips. The number of synthetic stars was 10\% of the observed sources for each of the frames and 30 of these modified images were obtained for each of the chips considered. PSF photometry was derived for these new images by applying the same PSF model used for the observed stars. We then estimated the fraction of synthetic stars recovered by our procedure for all the chips and a mean variation  of the completeness as a function of the magnitude was derived. In \figref{completitud} the percentage of recovered stars for the $g$ and $r$ bands is shown. Our DECam photometry recovers nearly all the synthetic stars up to $g, r \sim$ 22 and drops below 80\% at $g,r \sim$ 23.7. We set the limiting magnitude of our photometry at $g,r \sim$ 24.5, when the completeness reaches 50\%.    

\subsection{Radial density profile}
\label{secradial}

\begin{figure*}
     \begin{center}
      \includegraphics[scale=0.35]{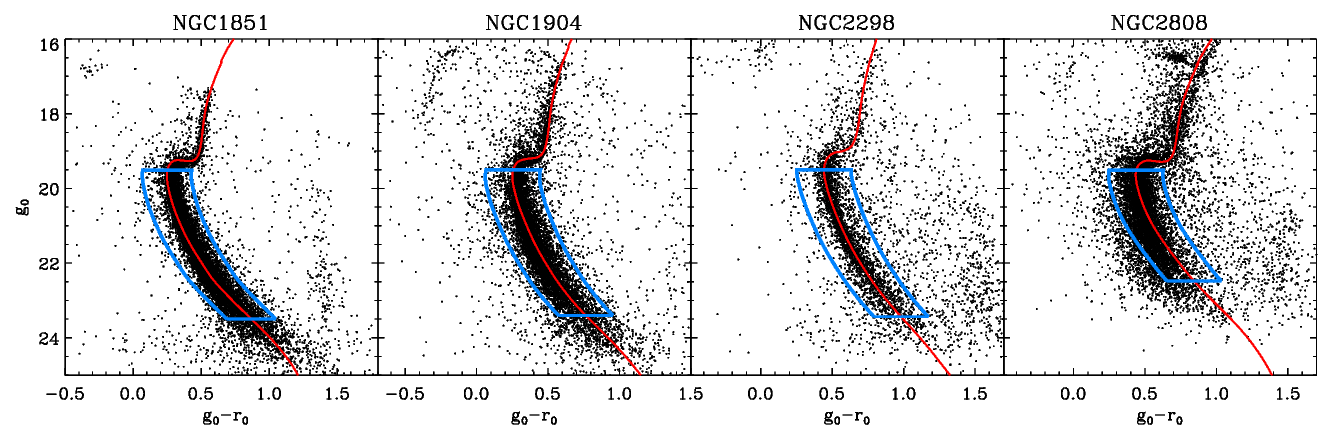}
      \includegraphics[scale=0.352]{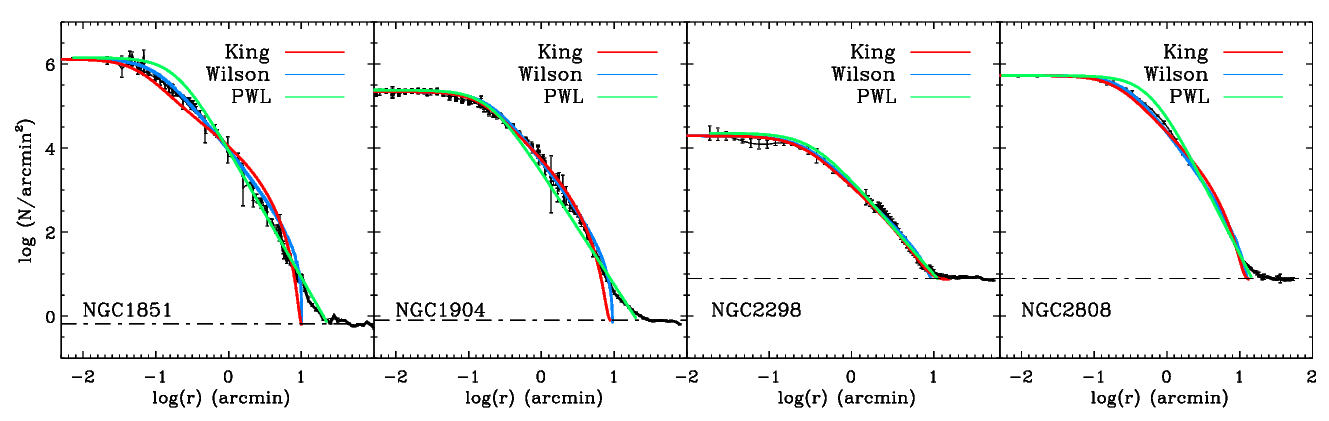}
      \caption[]{\emph{Top panels}: CMDs corresponding to the GCs (from left to right) NGC\,1851, NGC\,1904, NGC\,2298 and NGC\,2808. The solid red lines show the isochrones used to fit their CMDs and the blue boxes indicate the MS stars selected to generate the density radial profiles. \emph{Bottom panels}: radial density profiles obtained for the same GCs. The red, blue and green solid lines correspond to the best King, Wilson and power-law templates fitting, respectively. The dashed horizontal line represents the background density of stars located in the CMD selection box.}
\label{radialprofiles_plot}
     \end{center}
   \end{figure*}

We have derived the radial density profiles of the GCs by selecting cluster members along the MS. We first fitted  the CMD with a \cite{Dotter2008} isochrone with the corresponding parameters (age, metallicity) shown in \tabref{table1}, previously extracted from \cite{Harris2010} and \cite{Forbes2010}. We only considered stars with $|(g-r)_{\rm iso} - (g_{\rm 0}-r_{\rm 0})_{\rm CMD}|\, \leq\, 0.15$ in the magnitude range $23.5 > g_{\rm 0} > 19.5$ ($22.5 > g_{\rm 0} > 19.5$ for NGC\,2808), where $(g_{\rm 0}-r_{\rm 0})_{\rm iso}$ and $(g_{\rm 0}-r_{\rm 0})_{\rm CMD}$ are the isochrone and stars de-reddened colours, respectively (see \figref{radialprofiles_plot}). The adopted magnitude range represents a good compromise to maximize the number of cluster stars ensuring a completeness  
$\geq 90\%$.  Given that the presence of a small number of intruders in the sample (i.e. fore/background Galactic dwarf stars) is expected to be homogeneously distributed across the field of view, we assume that this contamination should not affect significantly the resulting profiles.

All the stars included in our selecting box were counted in concentric annuli, centered in the cluster coordinates (\tabref{table1}) with a fixed width in logarithmic scale. The number of counts per unit area has been obtained by dividing the number of stars  by the corresponding area covered by the annuli. The area of each annulus has been corrected for the presence of gaps between DECam chips and the field of view truncation and Poisson statistic for star counts has been considered to estimate the error on the density. Our exposure times were selected to detect outer GC stars but the innermost area of the clusters presents severe crowding problems. We thus complete our radial density profiles with the ones derived by \cite{Trager1995} for the globulars. A vertical scale factor was derived to match our data with their profiles by analyzing the spatial range in common between both datasets.

\subsection{Matched filter analysis}
\label{matched}

We generated two-dimensional density maps to search for tidal structures around the clusters and estimate their spatial extent. We performed a matched filter analysis as proposed by \cite{Rockosi2002} and widely used for the detection of extratidal features in Galactic GCs and halo substructures \citep[e.g.][]{Grillmair2006,Navarrete2017}. This technique makes use of more information about the CMD morphology of the cluster and background area than other methods based on the counting of MS stars. Although we refer the reader to the original reference for further details about this procedure, we briefly describe here the expressions implemented in this work.  

For a given spatial bin in which the field is divided, we assume that the density number of stars in every CMD bin in the diagram ($n_{\rm obs}$) is the result of a linear combination of the density of stars in the background ($n_{\rm back}$) and the density of cluster stars ($n_{\rm gc}$), in the following form

\begin{equation}
n_{\rm obs} = \omega\, n_{gc} + n_{back}.
\end{equation} 

Therefore, by minimizing the following expression and considering all the CMD bins, we derive the parameter $\omega$ which varies between 0 and 1, for spatial bins where no contribution to the observed CMD is coming from the GC or where only the cluster is generating the observed CMD, respectively: 

\begin{equation}
\chi^{2} = \sum^{g-r,g}_{i,j} \frac{(n_{\rm obs (i,j)} -[\omega\, n_{\rm gc (i,j)} +n_{\rm back (i,j)}])^{2}}{n_{\rm back (i,j)}}.
\end{equation}

The central arcminutes of the clusters are considered to derive $n_{\rm gc}$, while $n_{\rm back}$ results from the CMD corresponding to an area far away from these fields. Specifically, fields 2 and 6 are used to estimate the background CMD morphology for NGC\,1851 and NGC\,1904, respectively. For the clusters NGC\,2298 and NGC\,2808 the background was estimated from the region of constant density outside their tidal radii according to their radial profiles. The bin sizes used to compute $n_{\rm obs}$ and $n_{\rm back}$ were $\delta g_{\rm 0}=0.1$ and $\delta (g_{\rm 0}-r_{\rm 0})=$0.05. Using the isochrone corresponding to each cluster as reference, only those stars satisfying $|(g_{\rm 0}-r_{\rm 0})-(g-r)_{\rm iso}| < 0.15$ were considered for this procedure. The parameter $\omega$ is obtained for each of the spatial bins in which we divided the observed area and a density map is constructed for each of the clusters based on those results. To avoid false detections of structures produced by variations of the extinction in a wide field as the one provided by DECam, we derived $E(B-V)$ contours using the \cite{Schlafly2011} extinction maps. 

\subsection{Orbits}
\label{orbits}

In order to confirm if any of the features that might appear in our density maps are associated with the clusters, we have derived the orbits of the 4 GCs by adopting the \cite{Dinescu1999} and \cite{Dinescu2007} proper motion measurements and their radial velocities contained in the \cite{Harris2010} catalog. We have integrated their orbits backwards and forwards in a Milky Way-like potential considering the axisymmetric model of \cite{Allen1991}, which has a flattened disc, a bulge, and a spherical halo. The first two are modelled as Miyamoto--Nagai potentials \citep{Miyamoto1975}. The halo is a massive spherical component with the potential of Eq. 5 in \cite{Allen1991}. In this model, a value of $R_{\odot} = 8.5$\,kpc for the Sun's galactocentric distance and a circular speed of $V_{\rm c}(R_{\odot}) = 220$\,km\,s$^{-1}$ were adopted. We also adopt a Sun's motion of  $(U,V,W)$=(11, 12, 7)\,km\,s$^{-1}$ from \cite{Schonrich2010}. In \tabref{table2} we show the mean orbital parameters and their dispersions obtained after the orbit integration.

\begin{table*}
\centering
\begin{tabular}{lcclllcl}

Cluster & $R_{\rm G}$ (kpc)  & $\langle R_{\rm peri} \rangle$ (kpc) &  $\langle R_{\rm apo}\rangle$  (kpc)  &  $\langle Z_{\rm max}\rangle$  (kpc)  &  $\langle ecc\rangle$ & $\langle i \rangle$ (deg) & Orbital sense \\
\hline 
 NGC\,1851 & 16.6   & 6.2 $\pm$ 0.1&    37.0 $\pm$ 0.2&    12.7 $\pm$ 2.4&    0.71&      31$\pm$ 3&   prograde   \\ 
 NGC\,1904 & 18.8   & 4.4 $\pm$ 0.2&    21.4 $\pm$ 0.1&     7.8 $\pm$ 1.6&    0.66&      36$\pm$ 5&   prograde   \\ 
 NGC\,2298 & 15.8   & 2.7 $\pm$ 0.4&    17.6 $\pm$ 0.1&    11.5 $\pm$ 2.0&    0.73&      60$\pm$ 6&   retrograde \\
 NGC\,2808 & 11.1   & 2.9 $\pm$ 0.1&    12.8 $\pm$ 0.1&     3.7 $\pm$ 0.4&    0.63&      34$\pm$ 7&   prograde   \\ 
\hline
\end{tabular}
\caption[]{Orbital parameters derived for NGC\,1851, NGC\,1904, NGC\,2298 and NGC\,2808. The mean perigalactic and apogalactic distances, the mean maximum height above the Galactic plane, the mean orbital inclination (taken as 0$^{\circ}$ for completely planar orbits and 90$^{\circ}$ for polar orbits), the mean eccentricity, and the sense of the orbit are included.}
\label{table2}
\end{table*}

\section{Results and discussion}

  \begin{figure*}
     \begin{center}
      \includegraphics[scale=0.65]{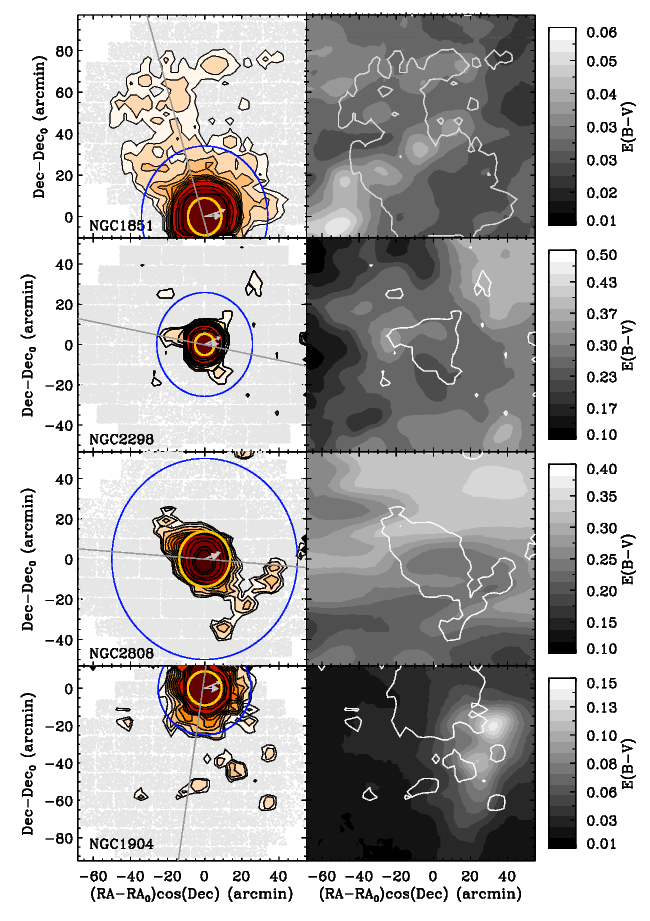}
      \caption[]{\emph{Left panels}: density maps corresponding to, from top to bottom, NGC\,1851, NGC\,2298, NGC\,2808 and NGC\,1904. The density contours correspond to the significance levels $\sigma =[1,1.5,2,3,4,5,6,7,8,10,50]$ above the background level. The yellow and blue solid lines indicate the King tidal and Jacobi radii for each cluster, respectively. The solid grey lines represent the tentative direction of the orbit of each cluster. The direction of the centre of the Galaxy and the Galactic plane are indicated by the big and small grey arrows, respectively. The footprints of the DECam fields have been included in the background. \emph{Right panels}: $E(B-V)$ maps from \cite{Schlafly2011} for the same area in the sky observed in fields \,1, 7, 11 and 12. The minimum density contours of the clusters contained in the left panels have been overplotted as a reference.}
\label{mapas_plot}
     \end{center}
   \end{figure*}

The CMDs and the radial density profiles corresponding to NGC\,1851, NGC\,1904, NGC\,2298 and NGC\,2808 are shown in \figref{radialprofiles_plot}. The radial density profiles were fitted by \cite{King1962}, \cite{Wilson1975}, and power-law \cite{Elson1987} models. The resulting parameters are contained in \tabref{table1} and the fitted templates are overplotted in the lower panel of \figref{radialprofiles_plot}. The density of objects within our selection boxes  remains nearly constant at large distances from the cluster centre. Therefore, we confirm that the contribution of fore/background stars and galaxies to our results is negligible, as we assumed in \secref{secradial} to construct the profiles.

\subsubsection*{NGC\,1851}

As already noted by \cite{Carballo-Bello2012}, power-law models are able to reproduce the outer section of the radial density profile of this cluster. As for the King model, the predicted distance of zero density level (tidal radius, $r_{\rm t}$) is smaller than the distance where our profile reachs the background level. This is crucial in order to define the term ``extra-tidal'' stars which is commonly used in studies focused on the search for stellar substructures around Galactic GCs. The obtained density profiles suggest that NGC\,1851 extends up to $>25$\,arcmin from its centre, where the background density is $\rho \sim 0.8$\,stars\,arcmin$^{-2}$.   

\begin{figure}
     \begin{center}
      \includegraphics[scale=0.35]{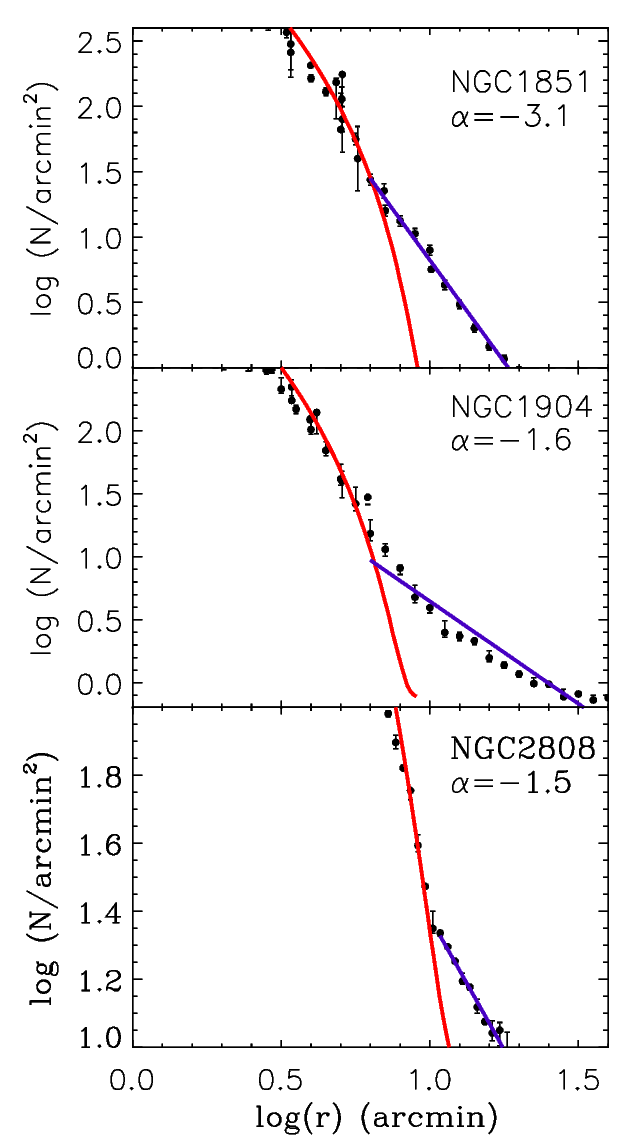}
      \caption[]{Outer radial density profiles for the GCs NGC\,1851, NGC\,1904 and NGC\,2808. The red solid line indicates the correspoding King model with the parameters shown in \tabref{table1}, while the blue solid line indicates the linear fit to the data beyond the break. The slope of the profile beyond the break is indicated by $\alpha$.}
\label{breaks}
     \end{center}
   \end{figure}

We identify a possible break (see \figref{breaks}) or change in the slope of the outer and less dense parts of the profile of NGC\,1851 \citep[also pointed out by][]{Grillmair1995}. The hypothetical break is observed at $r \sim 10$\,arcmin followed by a radial density profile falling as $r^{-3.1}$. This feature is usually generated by stars that either are lost by the cluster via tidal stripping or are still `potential escapers' trapped within the Jacobi radius, which corresponds to the first Lagrangian point \citep[see][]{Claydon2017,Daniel2017}. Clear examples of GCs with breaks in their profiles are NGC\,288 \citep{Leon2000}, Pal\,13 \citep{Cote2002}, Pal\,5 \citep{Odenkirchen2003}, NGC\,5466 \citep{Belokurov2006a}, AM\,4 \citep{Carraro2009}, Pal\,14 \citep{Sollima2011}, NGC\,6723 \citep{Chun2015}, NGC\,7089 \citep{Kuzma2016}, and, more recently, Eridanus and Pal\,15 \citep{Myeong2017}. The variation in the slope of the outer profile for NGC\,1851 might thus indicate that a similar process is also taking place in that cluster. However, not all GCs with breaks in the density profile display the characteristic `S-shaped' tidal tails.  

\cite{Leon2000} also found a break in the radial density profile of NGC\,1851 and unveiled tidal tails associated with this cluster and aligned with its orbital path. Interestingly, \cite{Olszewski2009} revealed a component of stars hypothetically associated with the cluster at distances greater than 1\,deg from its centre, which is equivalent to $\sim 250$\,pc at the distance of NGC\,1851. Similar structures have been observed around other Galactic GCs including NGC\,5694 \citep{Correnti2011} and NGC\,7089 \citep{Kuzma2016}. The intriguing NGC\,1851 halo of stars seems to fall as a power-law of $r^{-1.24}$ and would enclose a 0.1\% of the mass of the GC. More recently, \cite{Kuzma2017} confirmed the existence of an elliptical extended halo around the cluster in DECam data. That extended population of bound stars might be hard to understand in a cluster on a high-eccentricity orbit ($e \sim 0.7$, see \tabref{table2}) as that of NGC\,1851, which leads the cluster to cross the Galactic disc several times per Gyr. Different scenarios have been proposed to explain the survival of this diffuse stellar halo, including the formation of NGC\,1851 in the interior of a dwarf galaxy later accreted by the Milky Way and the merging of two individual GCs \citep{Carretta2011,Bekki2012}. In the numerical simulations generated by \cite{Bekki2012}, the hypothetical host dwarf galaxy of NGC\,1851 is almost completely destroyed in $\sim 10$\,Gyr and only a minor fraction of disc stars remains bound to the cluster (dwarf nucleus), which are the stars that compose the observed halo of stars around the globular. Those simulations also showed that the merging of GCs is possible in the interior of accreted dwarf galaxies ~-- even though, as pointed out by \cite{Catelan1997}, it would be unlikely for such a merger to involve two GCs with nearly the same metallicity, as would be required in this case \citep[see also][]{Amaro-Seoane2013}.

The density map obtained for NGC\,1851 is shown in \figref{mapas_plot}. The density contours exceed the King tidal radius (yellow circle) of the cluster and the isopleths with higher significance are observed as far out as $\sim 80$\,arcmin ($\sim 280$\,pc) from its centre, with no important extinction variations throughout the field of view. This result agrees with the detections of NGC\,1851 stars beyond $r \sim 250$\,pc by \cite{Olszewski2009} and \cite{Kuzma2017}. \cite{Baumgardt2010} estimated the Jacobi radius for the clusters included in this work following the \cite{Innanen1983} approximation:

\begin{equation}
r_{\rm J} = \left(\frac{G M_{\rm c}}{2 V_{\rm G}^{2}}\right)^{1/3} R_{\rm GC}^{\frac{2}{3}}.
\end{equation}

In this expression, $M_{\rm c}$ is the mass of the cluster, $V_{\rm G}$ the circular velocity of the galaxy and $R_{\rm GC}$ represents the Galactocentric distance of the globular. The Jacobi radius (blue circle in \figref{mapas_plot}) derived for NGC\,1851 is $r_{\rm J} \sim 34$\,arcmin and that volume appears to be almost completely populated by cluster stars. This result may suggest that this cluster is a clear candidate to be surrounded by lost stars in the form of tidal tails. The remarkable overdensity of stars beyond the Jacobi radius of NGC\,1851 revealed by our matched filter technique is located along the leading path of its orbit (as derived in \secref{orbits}, grey line). The position of this elongation has the same direction that the one found by \cite{Kuzma2017}. These evidence support the scenario in which NGC\,1851 has been partially disrupted by its interaction with the Milky Way. The observation of a larger area around this GC is necessary to confirm the morphology and real extension of these hypothetical tidal tails.

The outskirts of NGC\,1851 have also been explored with spectroscopy in recent years. \cite{Sollima2012} observed a sample of $\sim 107$ stars selected from the \cite{Carballo-Bello2014} catalogs and derived radial velocities. The radial velocity distribution obtained shows that an important number of stars with velocities compatible with that of NGC\,1851 are located beyond its King tidal radius. \cite{Marino2014} observed 23 stars at distances of up to $\sim 2.5\times r_{\rm t}$ and also obtained chemical abundances for a subsample of those targets. They found that most of these stars have radial velocities and metallicities compatible with that of NGC\,1851, and that they show Sr and Ba abundances resembling those of one of the sub-giant branch stars in this cluster. The latter would imply that one of the populations found in this GC is concentrated towards the centre of the cluster while the second (brighter sub-giant branch) one is found at larger distances \citep[see also][]{Milone2009,Zoccali2009}. New evidence by \cite{Navin2015} proved the presence of NGC\,1851 stars at $~3\times r_{\rm t}$, and \cite{Kunder2014} found a few stars in RAdial Velocity Experiment (RAVE) data \citep{Steinmetz2006} with a velocity similar to that of the GC and up to $\sim 10$\,deg away from the cluster centre. 

 \begin{figure}
     \begin{center}
      \includegraphics[scale=0.35]{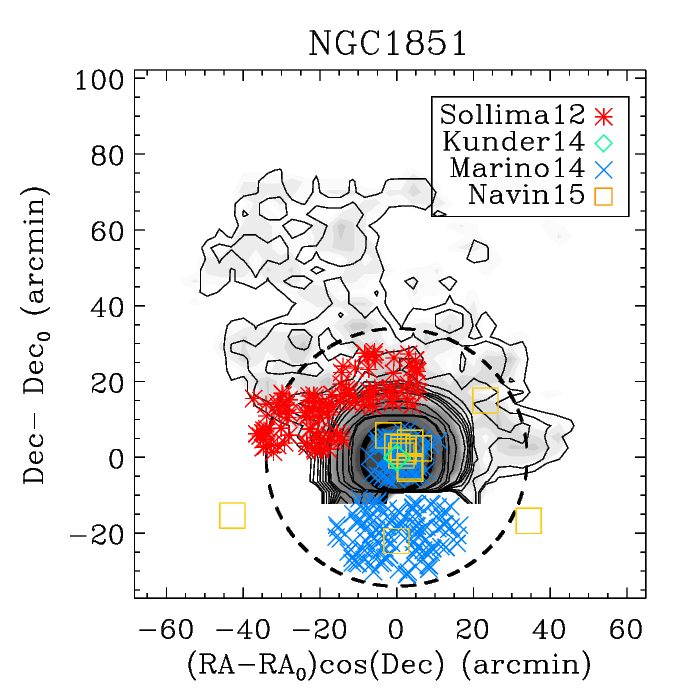}
      \caption[]{Density contours corresponding to NGC\,1851 with the same density levels displayed in \figref{mapas_plot}. The solid and dashed lines correspond, respectively, to the King tidal and Jacobi radii. The overplotted symbols correspond to spectroscopically targeted stars in and around NGC\,1851 from recent studies \citep{Sollima2012,Kunder2014,Marino2014,Navin2015}.}
\label{spectra_targets}
     \end{center}
   \end{figure}

\figref{spectra_targets} shows the position of outer NGC\,1851 stars observed by the above cited teams for the same section of the sky observed in field\,1. With the exception of a minor fraction of stars of Sollima's targets and 4 stars in the sample collected by \cite{Navin2015}, most of the spectroscopic target stars are contained between the Jacobi radius and  the King tidal radius. We have to note that NGC\,1851 lies in the edge of the corresponding DECam field, and so we are unable to establish density contours in the area from which \cite{Marino2014} selected most of their sample of stars. All these detections are probing an area of the sky well populated by NGC\,1851 as shown in \figref{spectra_targets}, so the resulting  radial velocity distributions mainly contain cluster stars. Therefore, the number of selected stars for spectroscopic follow-up seems to be insufficient and to be located too close to the cluster centre to study in detail the nature of the hypothetical halo of stars around NGC\,1851.

\subsubsection*{NGC\,1904}

A deviation from the King and Wilson models is found in the radial density profile of NGC\,1904, with a break at $r \sim 8$\,arcmin (see \figref{breaks}) followed by a profile falling as $r^{-1.6}$ and a background density level reached at $r > 25 $\,arcmin. In this case, the power-law template is also the best option to reproduce the outermost regions of the cluster. From the profiles and their CMDs, NGC\,1851 and NGC\,1904 might be considered \emph{twin} clusters since our results are remarkably similar for both globulars. 

\cite{Grillmair1995} systematically searched for tidal features around a sample of 12 GCs, including NGC\,1904, by applying a star count analysis on photographic photometry. Their density contours show that NGC\,1904 stars are found well beyond its King tidal radius and might generate the presence of a tentative southern tidal tail. This structure was not detected by \cite{Leon2000} but their results suggested the existence of a tail in the direction of the Galactic centre. \figref{mapas_plot} shows the density maps generated for this globular following the procedure described in \secref{matched}. As expected from its radial density profile and as suggested by Leon's results, the King tidal radius is once again exceeded by cluster stars and the volume defined by its Jacobi radius is almost completely filled. The overdensity unveiled by \cite{Leon2000} is not observed in \figref{mapas_plot} due to the position of the cluster close to the edge of the DECam field. 

Minor stellar overdensities observed in our maps might be identified as the tail detected by \cite{Grillmair1995} but seem to be coincident with extinction variations, so they may be artificial structures resulting from the matched filter analysis. On the other hand, these minor overdensities are not aligned with the trailing path of the orbit of NGC\,1904. The lack of a collimated structure within the DECam field of view as observed in other Galactic GCs with tidal tails makes it difficult to confirm their association with the cluster.

\subsubsection*{NGC\,2298}

Among the GCs included in this work, NGC\,2298 is the cluster where the outermost density profile is better fitted by all the models considered \citep[see][]{Carballo-Bello2012}. A low-mass cluster as NGC\,2298, on a very eccentric and retrograde orbit, may have easily lost stars because of its interaction with the Milky Way potential, and this should be reflected in its radial profile. Moreover, \cite{DeMarchi2007} studied the radial variation of the luminosity function and proved that this cluster only retains $\sim 15\%$ of its initial mass.  However, if NGC\,2298 was captured from an accreted galaxy, as suggested for all GCs on retrograde orbits, and therefore it was born in a lesser dense stellar system, this cluster may have remained \emph{tidally unaffected} as classified by \cite{Carballo-Bello2012}, and no evidence of breaks in the profile would be observed.  

The map for NGC\,2298 shown in \figref{mapas_plot} confirms that even when the cluster is well contained within its King tidal radius ($r_{\rm t} = 10.1$\,arcmin) as suggested by its radial profile, minor low-significance overdensities are unveiled around the cluster, predominantly in the volume defined by its Jacobi radius. \cite{Balbinot2011} used matched filter analysis to identify several stellar overdensities that might be associated with the tidal tails of NGC\,2298. Their density contours present a similar morphology to ours, although the DECam results contained in this work seem to identify those components as low-significance features. As also noted by \cite{Balbinot2011}, one of these elongations is orientated in the direction of the available orbit of the cluster, but the typical structure of a GC with tidal tails is not observed in our results.

  \begin{figure*}
     \begin{center}
      \includegraphics[scale=0.55]{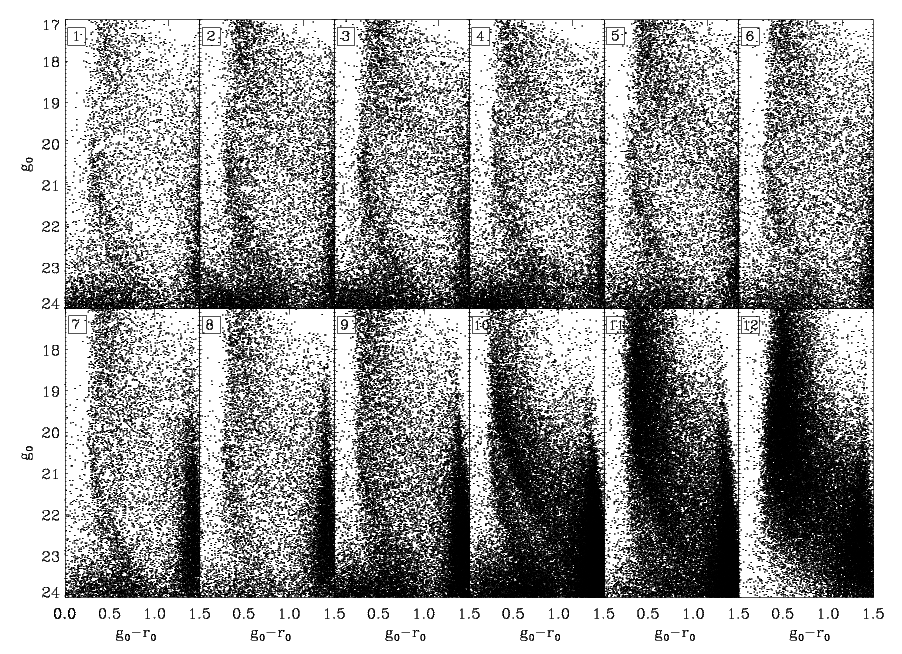}
      \caption[]{CMDs corresponding to the fields\,1--12. For the fields 1, 7, 11 and 12, we have included only those stars beyond 50\,arcmin from the centre of the clusters. The top left square indicates the field number.}
\label{todoscmds_plot}
     \end{center}
   \end{figure*}

\subsubsection*{NGC\,2808}

NGC\,2808 is the closest GC in our sample at only $d_{\odot} \sim 10$\,kpc. Its radial profile suggests that a small fraction of stars is located beyond $r \sim 10$\,arcmin where the density falls as $r^{-1.5}$. $N$-body simulations show that the radial distribution of potential escapers in the vicinity of a GC is steeper for clusters close to the perigalacticon \citep[e.g.][]{Kupper2010}. Indeed, the  profiles beyond the breaks for those globulars close to their apocentre positions (NGC\,2808 and NGC\,1904) are shallower than the one derived for NGC\,1851. Therefore, our results confirm that the orbital phase has a direct impact in the outer structure of Galactic GCs.

The density map confirms that significant overdensities are found beyond the King tidal radius and with a morphology resembling the classical `S-shaped' tidal tails observed around other Galactic GCs. There are not important extinction gradients coincident with the position of these overdensities. Interestingly, these elongations seem to be not aligned with the orbit that we have derived using \cite{Dinescu2007} proper motions. Long tidal tails as the ones observed in Pal\,5 and NGC\,5466 \citep[e.g.][]{Odenkirchen2001,Belokurov2006a,Grillmair2006} are usually considered good tracers for the orbits of those clusters. However, within the Jacobi radius, where stars are likely bound to the cluster, the orientation of the tails might reflect the Galaxy-GC interaction. Indeed, \cite{Montuori2007} N-body simulations suggest that while the orbital path of a GC is well described by the orientation of the tidal tails at large distances from the cluster centre, the emerging inner tails might be correlated with the orbital position of the cluster with respect to the Galactic centre \citep[see also][]{Lee2006a,Klimentowski2009}. For a cluster as NGC\,2808, close to its apocentre, their simulations predict inner tails aligned toward the Galactic centre. However, it is not the case of NGC\,2808, where the inner tails may aligned with the orbit that we have derived for the cluster. A larger field of view is necessary to explore the distribution and orientation of these tidal tails.

 \begin{figure}
     \begin{center}
      \includegraphics[scale=0.4]{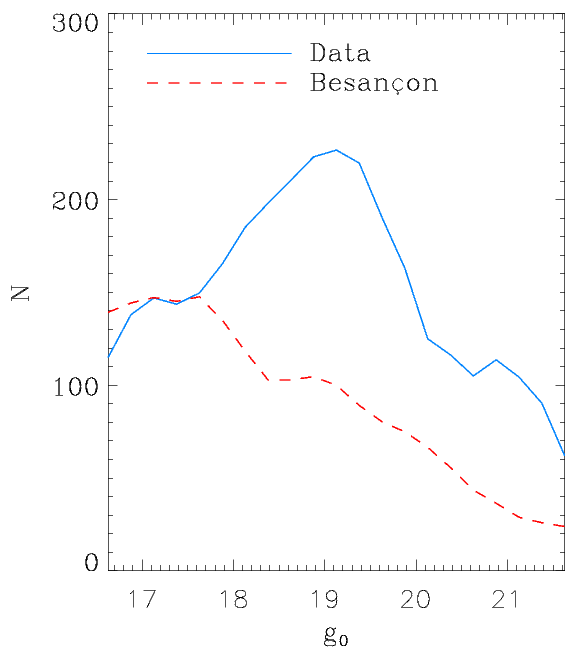}
      \caption[]{MS-TO star counts in the colour range $0.3 < g_{\rm 0}-r_{\rm 0} < 0.4$ for the CMD corresponding to the field\,10 (blue solid line) and for a synthetic diagram generated with the Besan\c con model for the same line of sight (red dahsed line). The red dahsed line has been scaled to fit the observations in the range $17 < g_{\rm 0} < 18$.}
\label{cuentas}
     \end{center}
   \end{figure}

\subsection{An underlying system around the clusters}

Accreted GCs are expected to be surrounded by the tidal remnants of their progenitor dwarf galaxies \citep[see][]{Carballo-Bello2014}. Once we have established the spatial extent of these GCs in the previous section, we proceed  to explore their surroundings searching for any hypothetical underlying systems. A hint of the existence of such a low-surface brightness population in the back/foreground of these clusters was reported by \cite{Sollima2012}. Their results unveiled a cold ($\sigma_{\rm v} \leq 20$\,km\,s$^{-1}$) peak at $v_{\rm r} \sim 180$\,km\,s$^{-1}$ in the radial velocity distribution for stars around NGC\,1851, consistent with the presence of a stream at the same heliocentric distance as that of the cluster. This component is not observed by any of the other spectroscopic studies around this cluster \citep[][]{Marino2014,Navin2015} nor by RAVE \citep{Kunder2014}, although the selection criteria of the latter are not comparable with Sollima's.
   
\figref{todoscmds_plot} contains the CMD corresponding to the 12 pointings observed with DECam for this work. For fields 1, 7, 11 and 12 we only included those stars located a distances greater than 50\,arcmin from the centre of the clusters. In a visual inspection, we observe an underlying MS in the fields 1--10 with $0.3 \lesssim g-r \lesssim 0.6$ and $g \gtrsim 20$. This component might be associated with the hypothetical underlying system that we are looking for around these clusters. That area of the CMD is not so  populated by stars in field 8 as in the rest of the fields with Galactic longitudes in the range $228^{\circ} < \ell < 243^{\circ}$. On the other hand, in field 10 we also find a parallel and brighter MS, which is also distinguishable in the CMD corresponding to field 11. As expected, the number of observed stars increases in the fields closer to the Galactic plane. 

We first investigate if the brighter MS is composed of Milky Way disc stars. We have compared our CMD for field\,10 with a synthetic diagram generated with the Besan\c con model for the same line of sight and solid angle \citep{Robin2003}. We defined a narrow colour range of 0.3 $< g_{\rm 0}-r_{\rm 0} < 0.4$ containing MS turn-off (MS TO) stars in both CMDs and counted the number of stars in each of the 30 $g_{\rm 0}-$magnitude bins in which the diagrams were divided. This procedure was repeated with 100 Besan\c con simulations and the results were averaged. The counts obtained are shown in \figref{cuentas}, where the sequence corresponding to the synthetic diagram has been shifted to match the counts derived for the bins in the range 17 $< g_{\rm 0} <$ 18. It is clear that the distribution of MS TO stars in the field\,10 shows a peak around $g_{\rm 0} \sim 19$ not observed in the one derived from Besan\c con. A secondary component is observed at $20 < g_{\rm 0} < 21 $, which seems to be associated with the fainter MS observed in most of our DECam fields. These results suggest that an unexpected stellar population is present along this line of sight.

Fields 10 and 11 are close to the position of the core of the hypothetically accreted Canis Major dwarf galaxy at ($\ell, b$) = ($240^{\circ},-8^{\circ}$) \citep{Martin2004}. The so-called ``blue-plume" of stars, a CMD feature classically associated with that overdensity, is not observed in our fields. This agrees with previous wide-field photometric surveys of the inner regions of Canis Major that show a concentration of those young blue stars towards the core of the accreted satellite \citep[e.g.][]{Butler2007}. The MS observed in fields 10 and 11 may correspond to the old MS population of Canis Major \citep[see example CMDs in][]{Butler2007}. Given that Canis Major might be connected with the Monoceros ring, we fitted a \cite{Dotter2008} isochrone by assuming ${\rm [Fe/H]} \sim -1$ and $age = 9$\,Gyr, as measured for Monoceros by different authors \citep[e.g.][]{Sollima2011,Carballo-Bello2015}. The distance derived for that population is $d_{\odot} = 7.6 \pm 1.5$\,kpc. which is similar to the heliocentric distance derived for Canis Major \citep[$d_{\odot} \sim 7.5$\,kpc][]{Martinez-Delgado2005,Butler2007,Bellazzini2006a}. This would indicate that at least NGC\,2298 (and likely NGC\,2808) is immersed in the Canis Major overdensity, as previously suggested by \cite{Martin2004}. 

The Monoceros ring, associated or not with the Canis Major overdensity, might also be reponsible for the underlying MS(s) found in our photometry. The predicted path of the Monoceros ring according to the \cite{Penarrubia2005} model is shown in \figref{monoceros}, on which the positions of our DECam fields are overplotted. All the fields observed for this work seem to be coincident in projected position with that of the Monoceros ring. Recent density maps generated with Pan-STARRS data show that the Monoceros ring is confined in the Galactic latitude range $-25^{\circ}  < b < +35^{\circ}$ \citep{Slater2014,Morganson2016}. Moreover, \cite{Morganson2016} found that this halo substructure might be composed of at least 2 concentric stellar rings centered at 4\,kpc from the Galaxy centre: one on the southern Galactic hemisphere at $d_{\odot} \sim 6$\,kpc and a second ring in the north with $d_{\odot} = 9$\,kpc \citep[see also][]{Conn2012}. Therefore, given the projected position and the distances derived, the MS observed exclusively in fields 10 and 11 may also be produced by the presence of stars associated with Monoceros.

  \begin{figure*}
     \begin{center}
      \includegraphics[scale=0.3]{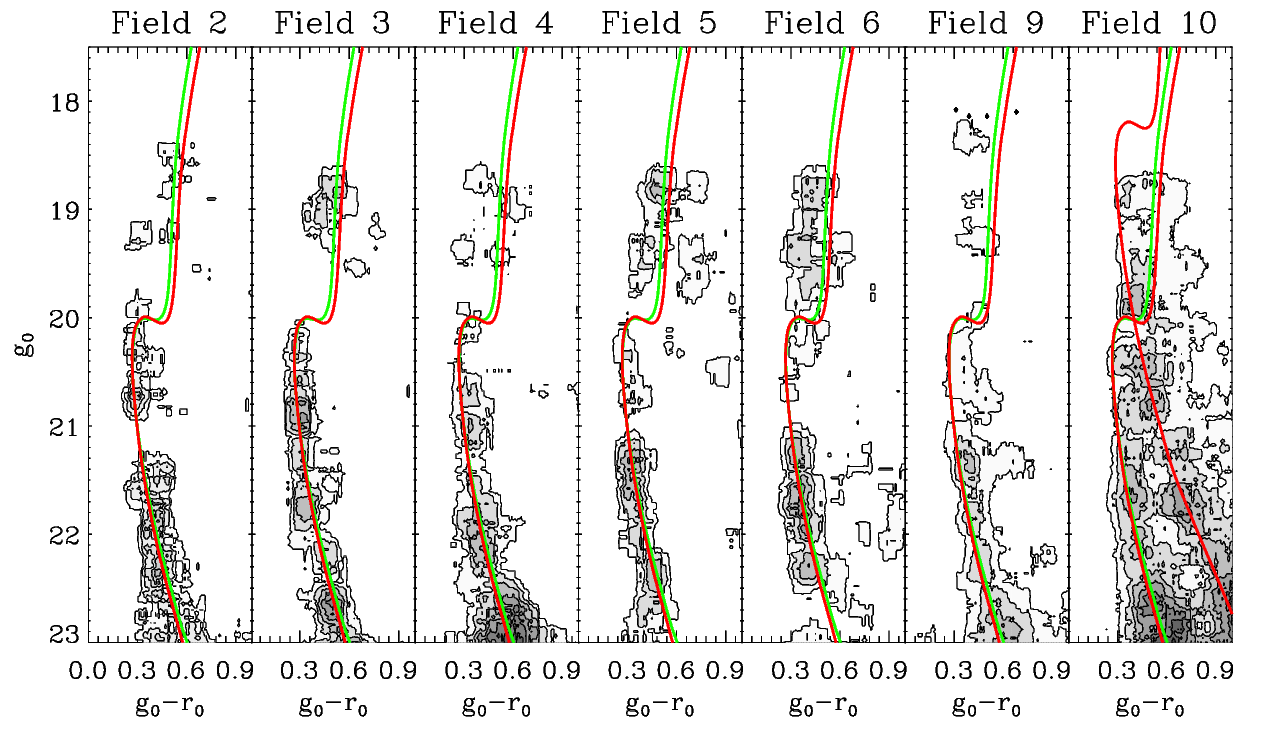}
      \caption[]{Hess diagram differences for the fields 2--6, 9 and 10 when the diagram corresponding to field 8 is substracted. The contours correspond to the significance levels $\sigma = [2,2.5,3,3.5]$. The red and green isochrones correspond to populations with $[Fe/H] = -1$ and  $age = 9$\,Gyr and $age = 12$\,Gyr, respectively.}
\label{hessdiagrams}
     \end{center}
   \end{figure*}

From the fields observed for this work and based on its projected distance with respect to the GCs, we have selected field 8 to sample the fore/background stellar populations. We have obtained a Hess diagram for each field by counting stars in the CMD with bin sizes of $\delta g = 0.1$\,mag and $\delta(g-r) = 0.01$\,mag. The resulting density maps were then smoothed with a boxcar average with a width of 5 bins. For all the fields observed we subtracted the Hess diagram derived for field 8. An underlying MS is found in all the CMDs corresponding to fields 1--11, as well as the bright sequences in fields 10 and 11 (see \figref{hessdiagrams}). The morphology and MS-TO magnitudes are nearly constant throughout the diagrams. These results suggest that a vast stellar substructure might be present around NGC\,1851 and NGC\,1904 (and possibly NGC\,2298) and spans at least 18\,deg $ \times 15$\,deg in the sky. \figref{hessdiagrams} shows the residual map for the fields 2 to 6 as examples of our results, including the fitted isochrones. 
 
The Monoceros ring described by \cite{Morganson2016} displays an important asymmetry north-south that might arise from the propagation of a density wave through the Galactic disc. The hypothesis of a Galactic plane modified by a stellar wave moving towards the outer regions of the Milky Way halo is also supported by results based on SDSS data by \cite{Xu2015}. Along the line of sight of the anticentre region ($110^{\circ} < \ell < 229^{\circ}$) and confined in $|b| < 15^{\circ}$, they found that Monoceros and other Galactic disc substructures \citep[e.g. the Triangulum Andromeda stream;][]{Rocha-Pinto2004,Majewski2004} might represent different sections of the same oscillating substructure with density peaks at $d_{\odot} \sim$ 10.5\,kpc (north), 12--14\,kpc (south), 16.5\,kpc (north), and 21\,kpc (south). The amplitude of the density oscillations associated with Monoceros in that scenario may be smaller than 1\,kpc in its inner region as derived by \cite{Xu2015} by fitting a simple model. More sophisticated numerical simulations also show that a vertical pattern might be caused by the impact of a dwarf galaxy similar to Sagittarius and disc stars may be \emph{kicked out} and put at larger $|Z|$ ranges \citep[e.g.]
[]{Purcell2011,Gomez2013,Price-Whelan2015}. More recently, \cite{Gomez2016} quantified the amplitude of the disc oscillations caused by a fly-by encounter with a satellite with $M \sim  4 \times 10^{10} M_{\odot}$ and found a $|Z|_{\rm max}$ of $\sim 2$\,kpc at $R > 15$\,kpc.  

In this context, we explore the nature of the second and more extended underlying population unveiled in our CMDs. When we fit the MS observed in most of the fields included in this work with the same Monoceros isochrone, we derive for that stellar population a heliocentric distance of $d_{\odot} = 18.2 \pm 2.0$\,kpc. Alternatively, we fitted isochrones with $age = 10$ and 12\,Gyr and [Fe/H] = -1 and derived distances of $d_{\odot} = 17.4 \pm 2.0$\,kpc and $15.5 \pm 2.0$\,kpc, respectively. This heliocentric distance range is compatible with the density peak observed by \cite{Xu2015} in the north and appears shifted only by a few kpc with respect to their south density peak at $d_{\odot} \sim 13$\,kpc in the anticentre region. At the mean Galactic latitude of the fields 1--7 ($b = -32.2^{\circ}$) and considering the heliocentric distances derived, we transform our values to the height below the Galactic plane and obtain -8.2\,kpc $< Z <$ -9.7\,kpc. We note that these values represent the mininum distance to the plane reached by the overdensity since it is not possible to determine its full extent with our data. If the stellar population unveiled in our photometry is associated with any of the hypothetical density waves moving towards the outer regions of the Milky Way, the amplitude would exceed the $Z_{\rm max}$ predicted by the numerical simulations mentioned above. On the other hand, the RR\,Lyrae to M giant stars ratio derived by \cite{Price-Whelan2015} for the stellar overdensities observed in the Triangulum-Andromeda region \citep[see also][]{Martin2007,Sheffield2014} suggests that those substructures might be composed of stars born in the Milky Way disc instead of an accreted dwarf galaxy. Interestingly, some of these overdensities revealed in photometric wide-sky surveys are found at $d_{\odot} \sim 15 - 20$\,kpc and at the same Galactic latitude as our fields, but with Galactic longitudes in the range $ 100^{\circ} < \ell < 160^{\circ}$.  Therefore, even when we are not able to prove or discard the presence of disc stars around the GCs NGC\,1851 and NGC\,1904, these results provide an opportunity to better constrain the orbit and initial mass of a satellite galaxy responsible for such a remarkable distortion in our Galaxy.

Besides the existence of Monoceros and/or the hypothetical density waves in the Galactic disc, there is enough observational evidence showing that the disc is warped and flared. The amplitude of the Milky Way disc warp has been estimated using different tracers. For example, \cite{Momany2006} selected red clump and red giant stars in Two Micron All Sky Survey (2MASS) to measure the amplitude of the warp and found a maximum at $\ell \sim 240^{\circ}$ with a mean deviation from the midplane of $\sim 3^{\circ}$. Moreover, \cite{Witham2008} studied the ditribution of H$\alpha$ emitters in the INT/WFC Photometric H$\alpha$ Survey (IPHAS) and found that for the Galactic longitude range $200^{\circ} < \ell < 300^{\circ}$ the mean latitude of H$\alpha$ emitters is contained in the range $-2^{\circ} < b < 0^{\circ}$. Therefore, the Galactic warp presents its maximum in a region of the sky with the same Galactic longitude range as the fields observed for this work, but its contribution would be relevant at higher latitudes than those of our fields. \cite{Lopez-Corredoira2014} proposed that the Milky Way flare ~-- radial variation of the scale height of the thick disc~-- exists beyond $R_{\rm G} = 15$\,kpc without truncation. Their fit to the star counts observed at low Galactic latitudes showed that the scale height of the disc might reach $h_{\rm z} \sim 4$\,kpc at 30\,kpc from the Milky Way centre. Given that the magnitude, extension and existence of the flare beyond $R_{\rm G} = 15$\,kpc and its connection with Monoceros are under debate, the presence of disc stars off-plane due to this mechanism along our line of sight is not completely discarded at this point, even when our stellar system is observed as far out as $Z \sim 10$\,kpc from the midplane.

In an alternative scenario, the component revealed in our data may be associated with the tentative stellar system connecting different Galactic halo overdensities  proposed by \cite{Li2016}. They discovered an overdensity of MS stars in DES data in the direction of the Eridanus and Phoenix constellations, centered at ($\ell,b) = (285^{\circ},-60^{\circ})$ and lying at $d_{\odot} \sim 16$\,kpc. When a polar orbit at $\sim 18$\,kpc from the centre of the Milky Way and crossing the Eridanus-Phoenix overdensity is considered, they found that both the Virgo overdensity \citep[e.g.][$d_{\odot} \sim 6-20$\,kpc]{Juric2008,Duffau2014,Vivas2016} and the Hercules-Aquila cloud \citep[][$d_{\odot} \sim 10-20$\,kpc]{Belokurov2007a,Simion2014} are also coincident in projected position. Moreover, such a hypothetical vast plane of connected stellar overdensities is close to the plane of Milky Way satellites and GCs proposed by \cite{Pawlowski2012}. However, the polar orbit proposed by \cite{Li2016} crosses the Galactic plane at $\ell \sim 260^{\circ}$, while all our fields are located at $\ell < 245^{\circ}$. Even so, this scenario is worth of future exploration, given that the exact path of that hypothetical polar orbit is unclear and the heliocentric distance range is compatible with the underlying stellar population found around our GCs. 

The final and most exciting possibility is that the system around these GCs  represents a new stellar Galactic halo overdensity covering a wide area of the sky. As we have already commented above, these GCs (specially NGC\,1851) have been considered potential accreted clusters because of their individual properties and as a \emph{cluster of clusters}. However, the association of all these clusters with the same hypothetical accreted system  \citep[as suggested in][]{Bellazzini2003} may be difficult to reconcile with the orbits derived for these clusters in this work, with NGC\,2298 on a retrograde orbit around the Galaxy. On the other hand, NGC\,1851, NGC\,1904 and NGC\,2808 have similar mean orbital inclinations and the same sense of rotation. The stellar system unveiled in our data lies at 15.5\,kpc$ < d_{\odot} < 18.2$\,kpc while NGC\,1851, NGC\,1904, NGC\,2298 and NGC\,2808 are found at $d_{\odot} = 12.1, 12.9, 10.8$ and 9.6\,kpc, respectively, with a mean uncertainty of $<\sigma_{d}> \sim 2$\,kpc \citep[see references in][]{Harris2010}. More recently, \cite{Wagner-Kaiser2017} have derived new distances for NGC\,1851, NGC\,2298 and NGC\,2808 and found that these clusters lie in the narrow distance range $15.53 < d_{\odot} < 15.65$\,kpc. Therefore, both NGC\,1851 and NGC\,1904 have heliocentric distances compatible, within the errors, with the location of this stellar overdensity. Our results might indicate that both globulars are immersed in a Galactic halo stellar substructure, which may correspond to the tidal remnants of their progenitor galaxy. However, a spatial coincidence of a GC and a tidal stream or stellar substructure is not the only way to identify extra-Galactic clusters. Follow-up spectroscopy, as recently obtained by \cite{Carballo-Bello2017} for a sample of stars around Whiting\,1, is required to establish the nature of the stellar system revealed in our DECam photometry and confirm its hypothetical association with NGC\,1851 and NGC\,1904.

\section{Conclusions}

We have observed 12 fields with DECam containing the GCs NGC\,1851, NGC\,1904, NGC\,2298 and NGC\,2808 and their surroundings. The derived radial density profiles and the density maps generated using the matched filter technique suggest that an important fraction of cluster stars are found beyond their King tidal radii. Our results show that all the clusters display structures within their Jacobi radii, with the exception of NGC\,1851, where a tail seems to emerge from the cluster, parallel to the orbit derived for this cluster in this work.

When the Milky Way components are subtracted from the obtained CMDs using a control field at similar Galactic latitude, the presence of an underliying system is evident in almost all the fields included in this work, together with a second component only observed at higher latitudes. The latter is compatible in distance and projected position with the Canis Major overdensity and the Monoceros ring, a   controversial halo substructure composed of different non-symmetric stellar rings. The more extended population, found in an area of $18\,{\rm deg} \times 15\,{\rm deg}$ in the sky, is found at $<d_{\odot}> \sim 16$\,kpc and reaching a distance to the plane of $Z \sim 10$\,kpc.  

We propose different scenarios to explain the presence of this stellar system in the anticentre direction: a section of the Monoceros ring, when this substructure is generated by a density wave travelling through the disc towards the outer regions of the Galaxy; a combined effect between Monoceros and the Galactic flared disc; a detection of the hypothetical stream connecting the Eridanus-Phonenix and Virgo overdensities; or the tidal remnants of the tentative progenitor dwarf galaxy host of NGC\,1851 and NGC\,1904. Only with follow-up spectroscopy will it be possible to explain our detections in terms of one (or a mix) of these hypotheses.

\section*{Acknowledgements}

JAC-B acknowledges financial support to CONICYT-Chile FONDECYT Postdoctoral Fellowship 3160502. JAC-B, CN and MC received support from the Ministry for the Economy, Development, and Tourism's Programa Iniciativa Cient\'ifica Milenio through grant IC120009, awarded to the Millennium Institute of Astrophysics (MAS), and from CONICYT's PCI program through grant DPI20140066. CN acknowledges support from CONICYT-PCHA/Doctorado Nacional 2015-21151643. DMD acknowledges support by Sonderforschungsbereich (SFB) 881 ``The Milky Way System" of the German Research Foundation (DFG), particularly through subproject A2. MC acknowledges additional support by FONDECYT Project 1171273 and the Basal-CONICYT Center
for Astrophysics and Associated Technologies (PFB-06). R. R. M. acknowledges partial support from project Basal PFB-$06$ as well as FONDECYT project N$^{\circ}117064$. This project used data obtained with the Dark Energy Camera (DECam), which was constructed by the Dark Energy Survey (DES) collaboration. Funding for the DES Projects has been provided by 
the U.S. Department of Energy, 
the U.S. National Science Foundation, 
the Ministry of Science and Education of Spain, 
the Science and Technology Facilities Council of the United Kingdom, 
the Higher Education Funding Council for England, 
the National Center for Supercomputing Applications at the University of Illinois at Urbana-Champaign, 
the Kavli Institute of Cosmological Physics at the University of Chicago, 
the Center for Cosmology and Astro-Particle Physics at the Ohio State University, 
the Mitchell Institute for Fundamental Physics and Astronomy at Texas A\&M University, 
Financiadora de Estudos e Projetos, Funda{\c c}{\~a}o Carlos Chagas Filho de Amparo {\`a} Pesquisa do Estado do Rio de Janeiro, 
Conselho Nacional de Desenvolvimento Cient{\'i}fico e Tecnol{\'o}gico and the Minist{\'e}rio da Ci{\^e}ncia, Tecnologia e Inovac{\~a}o, 
the Deutsche Forschungsgemeinschaft, 
and the Collaborating Institutions in the Dark Energy Survey. 
The Collaborating Institutions are 
Argonne National Laboratory, 
the University of California at Santa Cruz, 
the University of Cambridge, 
Centro de Investigaciones En{\'e}rgeticas, Medioambientales y Tecnol{\'o}gicas-Madrid, 
the University of Chicago, 
University College London, 
the DES-Brazil Consortium, 
the University of Edinburgh, 
the Eidgen{\"o}ssische Technische Hoch\-schule (ETH) Z{\"u}rich, 
Fermi National Accelerator Laboratory, 
the University of Illinois at Urbana-Champaign, 
the Institut de Ci{\`e}ncies de l'Espai (IEEC/CSIC), 
the Institut de F{\'i}sica d'Altes Energies, 
Lawrence Berkeley National Laboratory, 
the Ludwig-Maximilians Universit{\"a}t M{\"u}nchen and the associated Excellence Cluster Universe, 
the University of Michigan, 
{the} National Optical Astronomy Observatory, 
the University of Nottingham, 
the Ohio State University, 
the University of Pennsylvania, 
the University of Portsmouth, 
SLAC National Accelerator Laboratory, 
Stanford University, 
the University of Sussex, 
and Texas A\&M University.

\def\jnl@style{\it}                       
\def\mnref@jnl#1{{\jnl@style#1}}          
\def\aj{\mnref@jnl{AJ}}                   
\def\apj{\mnref@jnl{ApJ}}                 
\def\apjl{\mnref@jnl{ApJL}}               
\def\aap{\mnref@jnl{A\&A}}                
\def\mnras{\mnref@jnl{MNRAS}}             
\def\nat{\mnref@jnl{Nat.}}                
\def\iaucirc{\mnref@jnl{IAU~Circ.}}       
\def\atel{\mnref@jnl{ATel}}               
\def\iausymp{\mnref@jnl{IAU~Symp.}}       
\def\pasp{\mnref@jnl{PASP}}               
\def\araa{\mnref@jnl{ARA\&A}}             
\def\apjs{\mnref@jnl{ApJS}}               
\def\aapr{\mnref@jnl{A\&A Rev.}}          

\bibliographystyle{mn2e}
\bibliography{biblio}

\begin{thebibliography}{}

\bibitem[\protect\citeauthoryear{{Allen} \& {Santillan}}{{Allen} \&
  {Santillan}}{1991}]{Allen1991}
{Allen} C.,  {Santillan} A.,  1991, \rmxaa, 22, 255

\bibitem[\protect\citeauthoryear{{Amaro-Seoane}, {Konstantinidis}, {Brem} \&
  {Catelan}}{{Amaro-Seoane} et~al.}{2013}]{Amaro-Seoane2013}
{Amaro-Seoane} P.,  {Konstantinidis} S.,  {Brem} P.,    {Catelan} M.,  2013,
  \mnras, 435, 809

\bibitem[\protect\citeauthoryear{{Balbinot}, {Santiago}, {da Costa}, {Makler}
  \& {Maia}}{{Balbinot} et~al.}{2011}]{Balbinot2011}
{Balbinot} E.,  {Santiago} B.~X.,  {da Costa} L.~N.,  {Makler} M.,    {Maia}
  M.~A.~G.,  2011, \mnras, 416, 393

\bibitem[\protect\citeauthoryear{{Baumgardt}, {Parmentier}, {Gieles} \&
  {Vesperini}}{{Baumgardt} et~al.}{2010}]{Baumgardt2010}
{Baumgardt} H.,  {Parmentier} G.,  {Gieles} M.,    {Vesperini} E.,  2010,
  \mnras, 401, 1832

\bibitem[\protect\citeauthoryear{{Bekki} \& {Yong}}{{Bekki} \&
  {Yong}}{2012}]{Bekki2012}
{Bekki} K.,  {Yong} D.,  2012, \mnras, 419, 2063

\bibitem[\protect\citeauthoryear{{Bellazzini}, {Ferraro} \&
  {Ibata}}{{Bellazzini} et~al.}{2002}]{Bellazzini2002}
{Bellazzini} M.,  {Ferraro} F.~R.,    {Ibata} R.,  2002, \aj, 124, 915

\bibitem[\protect\citeauthoryear{{Bellazzini}, {Ferraro} \&
  {Ibata}}{{Bellazzini} et~al.}{2003}]{Bellazzini2003}
{Bellazzini} M.,  {Ferraro} F.~R.,    {Ibata} R.,  2003, \aj, 125, 188

\bibitem[\protect\citeauthoryear{{Bellazzini}, {Ibata} \&
  {Ferraro}}{{Bellazzini} et~al.}{2004}]{Bellazzini2004}
{Bellazzini} M.,  {Ibata} R.,    {Ferraro} F.~R.,  2004, in {F.~Prada,
  D.~Martinez Delgado, \& T.~J.~Mahoney} ed., Satellites and Tidal Streams
  Vol.~327 of Astronomical Society of the Pacific Conference Series, {Globular
  Clusters in the Sgr Stream and Other Structures}.
pp 220--+

\bibitem[\protect\citeauthoryear{{Bellazzini}, {Ibata}, {Martin}, {Lewis},
  {Conn} \& {Irwin}}{{Bellazzini} et~al.}{2006}]{Bellazzini2006a}
{Bellazzini} M.,  {Ibata} R.,  {Martin} N.,  {Lewis} G.~F.,  {Conn} B.,
  {Irwin} M.~J.,  2006, \mnras, 366, 865

\bibitem[\protect\citeauthoryear{{Belokurov}, {Evans}, {Irwin}, {Hewett} \&
  {Wilkinson}}{{Belokurov} et~al.}{2006}]{Belokurov2006a}
{Belokurov} V.,  {Evans} N.~W.,  {Irwin} M.~J.,  {Hewett} P.~C.,    {Wilkinson}
  M.~I.,  2006, \apjl, 637, L29

\bibitem[\protect\citeauthoryear{{Belokurov et al.}}{{Belokurov et
  al.}}{2006}]{Belokurov2006}
{Belokurov et al.} 2006, \apjl, 642, L137

\bibitem[\protect\citeauthoryear{{Belokurov et al.}}{{Belokurov et
  al.}}{2007}]{Belokurov2007a}
{Belokurov et al.} 2007, \apjl, 657, L89

\bibitem[\protect\citeauthoryear{{Butler}, {Mart{\'{\i}}nez-Delgado}, {Rix},
  {Pe{\~n}arrubia} \& {de Jong}}{{Butler} et~al.}{2007}]{Butler2007}
{Butler} D.~J.,  {Mart{\'{\i}}nez-Delgado} D.,  {Rix} H.-W.,  {Pe{\~n}arrubia}
  J.,    {de Jong} J.~T.~A.,  2007, \aj, 133, 2274

\bibitem[\protect\citeauthoryear{{Carballo-Bello}, {Gieles}, {Sollima},
  {Koposov}, {Mart{\'{\i}}nez-Delgado} \& {Pe{\~n}arrubia}}{{Carballo-Bello}
  et~al.}{2012}]{Carballo-Bello2012}
{Carballo-Bello} J.~A.,  {Gieles} M.,  {Sollima} A.,  {Koposov} S.,
  {Mart{\'{\i}}nez-Delgado} D.,    {Pe{\~n}arrubia} J.,  2012, \mnras, 419, 14

\bibitem[\protect\citeauthoryear{{Carballo-Bello}, {Mu{\~n}oz}, {Carlin},
  {C{\^o}t{\'e}}, {Geha}, {Simon} \& {Djorgovski}}{{Carballo-Bello}
  et~al.}{2015}]{Carballo-Bello2015}
{Carballo-Bello} J.~A.,  {Mu{\~n}oz} R.~R.,  {Carlin} J.~L.,  {C{\^o}t{\'e}}
  P.,  {Geha} M.,  {Simon} J.~D.,    {Djorgovski} S.~G.,  2015, \apj, 805, 51

\bibitem[\protect\citeauthoryear{{Carballo-Bello}, {Sollima},
  {Mart{\'{\i}}nez-Delgado}, {Pila-D{\'{\i}}ez}, {Leaman}, {Fliri}, {Mu{\~n}oz}
  \& {Corral-Santana}}{{Carballo-Bello} et~al.}{2014}]{Carballo-Bello2014}
{Carballo-Bello} J.~A.,  {Sollima} A.,  {Mart{\'{\i}}nez-Delgado} D.,
  {Pila-D{\'{\i}}ez} B.,  {Leaman} R.,  {Fliri} J.,  {Mu{\~n}oz} R.~R.,
  {Corral-Santana} J.~M.,  2014, \mnras, 445, 2971

\bibitem[\protect\citeauthoryear{{Carballo-Bello et al.}}{{Carballo-Bello et
  al.}}{2017}]{Carballo-Bello2017}
{Carballo-Bello et al.} 2017, \mnras, 467, L91

\bibitem[\protect\citeauthoryear{{Carraro}}{{Carraro}}{2009}]{Carraro2009}
{Carraro} G.,  2009, \aj, 137, 3809

\bibitem[\protect\citeauthoryear{{Carretta}, {Lucatello}, {Gratton},
  {Bragaglia} \& {D'Orazi}}{{Carretta} et~al.}{2011}]{Carretta2011}
{Carretta} E.,  {Lucatello} S.,  {Gratton} R.~G.,  {Bragaglia} A.,    {D'Orazi}
  V.,  2011, \aap, 533, A69

\bibitem[\protect\citeauthoryear{{Casetti-Dinescu}, {Girard}, {Herrera}, {van
  Altena}, {L{\'o}pez} \& {Castillo}}{{Casetti-Dinescu}
  et~al.}{2007}]{Dinescu2007}
{Casetti-Dinescu} D.~I.,  {Girard} T.~M.,  {Herrera} D.,  {van Altena} W.~F.,
  {L{\'o}pez} C.~E.,    {Castillo} D.~J.,  2007, \aj, 134, 195

\bibitem[\protect\citeauthoryear{{Catelan}}{{Catelan}}{1997}]{Catelan1997}
{Catelan} M.,  1997, \apjl, 478, L99

\bibitem[\protect\citeauthoryear{{Chambers et al.}}{{Chambers et
  al.}}{2016}]{Chambers2016}
{Chambers et al.} 2016, ArXiv e-prints

\bibitem[\protect\citeauthoryear{{Chun}, {Kang}, {Jung} \& {Sohn}}{{Chun}
  et~al.}{2015}]{Chun2015}
{Chun} S.-H.,  {Kang} M.,  {Jung} D.,    {Sohn} Y.-J.,  2015, \aj, 149, 29

\bibitem[\protect\citeauthoryear{{Claydon}, {Gieles} \& {Zocchi}}{{Claydon}
  et~al.}{2017}]{Claydon2017}
{Claydon} I.,  {Gieles} M.,    {Zocchi} A.,  2017, \mnras, 466, 3937

\bibitem[\protect\citeauthoryear{{Conn}, {Lewis}, {Irwin}, {Ibata}, {Ferguson},
  {Tanvir} \& {Irwin}}{{Conn} et~al.}{2005}]{Conn2005}
{Conn} B.~C.,  {Lewis} G.~F.,  {Irwin} M.~J.,  {Ibata} R.~A.,  {Ferguson}
  A.~M.~N.,  {Tanvir} N.,    {Irwin} J.~M.,  2005, \mnras, 362, 475

\bibitem[\protect\citeauthoryear{{Conn}, {No{\"e}l}, {Rix}, {Lane}, {Lewis},
  {Irwin}, {Martin}, {Ibata}, {Dolphin} \& {Chapman}}{{Conn}
  et~al.}{2012}]{Conn2012}
{Conn} B.~C.,  {No{\"e}l} N.~E.~D.,  {Rix} H.-W.,  {Lane} R.~R.,  {Lewis}
  G.~F.,  {Irwin} M.~J.,  {Martin} N.~F.,  {Ibata} R.~A.,  {Dolphin} A.,
  {Chapman} S.,  2012, \apj, 754, 101

\bibitem[\protect\citeauthoryear{{Correnti}, {Bellazzini}, {Dalessandro},
  {Mucciarelli}, {Monaco} \& {Catelan}}{{Correnti} et~al.}{2011}]{Correnti2011}
{Correnti} M.,  {Bellazzini} M.,  {Dalessandro} E.,  {Mucciarelli} A.,
  {Monaco} L.,    {Catelan} M.,  2011, \mnras, 417, 2411

\bibitem[\protect\citeauthoryear{{C{\^o}t{\'e}}, {Djorgovski}, {Meylan},
  {Castro} \& {McCarthy}}{{C{\^o}t{\'e}} et~al.}{2002}]{Cote2002}
{C{\^o}t{\'e}} P.,  {Djorgovski} S.~G.,  {Meylan} G.,  {Castro} S.,
  {McCarthy} J.~K.,  2002, \apj, 574, 783

\bibitem[\protect\citeauthoryear{{Cummings}, {Geisler}, {Villanova} \&
  {Carraro}}{{Cummings} et~al.}{2014}]{Cummings2014}
{Cummings} J.~D.,  {Geisler} D.,  {Villanova} S.,    {Carraro} G.,  2014, \aj,
  148, 27

\bibitem[\protect\citeauthoryear{{Daniel}, {Heggie} \& {Varri}}{{Daniel}
  et~al.}{2017}]{Daniel2017}
{Daniel} K.~J.,  {Heggie} D.~C.,    {Varri} A.~L.,  2017, \mnras, 468, 1453

\bibitem[\protect\citeauthoryear{{D'Antona}, {Bellazzini}, {Caloi}, {Pecci},
  {Galleti} \& {Rood}}{{D'Antona} et~al.}{2005}]{Dantona2005}
{D'Antona} F.,  {Bellazzini} M.,  {Caloi} V.,  {Pecci} F.~F.,  {Galleti} S.,
  {Rood} R.~T.,  2005, \apj, 631, 868

\bibitem[\protect\citeauthoryear{{Dark Energy Survey Collaboration}}{{Dark
  Energy Survey Collaboration}}{2016}]{DES2016}
{Dark Energy Survey Collaboration} 2016, \mnras, 460, 1270

\bibitem[\protect\citeauthoryear{{de Marchi} \& {Pulone}}{{de Marchi} \&
  {Pulone}}{2007}]{DeMarchi2007}
{de Marchi} G.,  {Pulone} L.,  2007, \aap, 467, 107

\bibitem[\protect\citeauthoryear{{Dinescu}, {Girard} \& {van Altena}}{{Dinescu}
  et~al.}{1999}]{Dinescu1999}
{Dinescu} D.~I.,  {Girard} T.~M.,    {van Altena} W.~F.,  1999, \aj, 117, 1792

\bibitem[\protect\citeauthoryear{{D'Orazi}, {Gratton}, {Angelou}, {Bragaglia},
  {Carretta}, {Lattanzio}, {Lucatello}, {Momany}, {Sollima} \&
  {Beccari}}{{D'Orazi} et~al.}{2015}]{Orazi2015}
{D'Orazi} V.,  {Gratton} R.~G.,  {Angelou} G.~C.,  {Bragaglia} A.,  {Carretta}
  E.,  {Lattanzio} J.~C.,  {Lucatello} S.,  {Momany} Y.,  {Sollima} A.,
  {Beccari} G.,  2015, \mnras, 449, 4038

\bibitem[\protect\citeauthoryear{{Dotter}, {Chaboyer}, {Jevremovi{\'c}},
  {Kostov}, {Baron} \& {Ferguson}}{{Dotter} et~al.}{2008}]{Dotter2008}
{Dotter} A.,  {Chaboyer} B.,  {Jevremovi{\'c}} D.,  {Kostov} V.,  {Baron} E.,
   {Ferguson} J.~W.,  2008, \apjs, 178, 89

\bibitem[\protect\citeauthoryear{{Duffau}, {Vivas}, {Zinn}, {M{\'e}ndez} \&
  {Ruiz}}{{Duffau} et~al.}{2014}]{Duffau2014}
{Duffau} S.,  {Vivas} A.~K.,  {Zinn} R.,  {M{\'e}ndez} R.~A.,    {Ruiz} M.~T.,
  2014, \aap, 566, A118

\bibitem[\protect\citeauthoryear{{Elson}, {Fall} \& {Freeman}}{{Elson}
  et~al.}{1987}]{Elson1987}
{Elson} R.~A.~W.,  {Fall} S.~M.,    {Freeman} K.~C.,  1987, \apj, 323, 54

\bibitem[\protect\citeauthoryear{{Erkal} \& {Belokurov}}{{Erkal} \&
  {Belokurov}}{2015}]{Erkal2015}
{Erkal} D.,  {Belokurov} V.,  2015, \mnras, 450, 1136

\bibitem[\protect\citeauthoryear{{Flaugher et al.}}{{Flaugher et
  al.}}{2015}]{Flaugher2015}
{Flaugher et al.} 2015, \aj, 150, 150

\bibitem[\protect\citeauthoryear{{Font}, {Benson}, {Bower}, {Frenk}, {Cooper},
  {De Lucia}, {Helly}, {Helmi}, {Li}, {McCarthy}, {Navarro}, {Springel},
  {Starkenburg}, {Wang} \& {White}}{{Font} et~al.}{2011}]{Font2011}
{Font} A.~S.,  {Benson} A.~J.,  {Bower} R.~G.,  {Frenk} C.~S.,  {Cooper} A.,
  {De Lucia} G.,  {Helly} J.~C.,  {Helmi} A.,  {Li} Y.-S.,  {McCarthy} I.~G.,
  {Navarro} J.~F.,  {Springel} V.,  {Starkenburg} E.,  {Wang} J.,    {White}
  S.~D.~M.,  2011, \mnras, 417, 1260

\bibitem[\protect\citeauthoryear{{Forbes} \& {Bridges}}{{Forbes} \&
  {Bridges}}{2010}]{Forbes2010}
{Forbes} D.~A.,  {Bridges} T.,  2010, \mnras, 404, 1203

\bibitem[\protect\citeauthoryear{{G{\'o}mez}, {Helmi}, {Cooper}, {Frenk},
  {Navarro} \& {White}}{{G{\'o}mez} et~al.}{2013}]{Gomez2013}
{G{\'o}mez} F.~A.,  {Helmi} A.,  {Cooper} A.~P.,  {Frenk} C.~S.,  {Navarro}
  J.~F.,    {White} S.~D.~M.,  2013, \mnras, 436, 3602

\bibitem[\protect\citeauthoryear{{G{\'o}mez}, {White}, {Marinacci}, {Slater},
  {Grand}, {Springel} \& {Pakmor}}{{G{\'o}mez} et~al.}{2016}]{Gomez2016}
{G{\'o}mez} F.~A.,  {White} S.~D.~M.,  {Marinacci} F.,  {Slater} C.~T.,
  {Grand} R.~J.~J.,  {Springel} V.,    {Pakmor} R.,  2016, \mnras, 456, 2779

\bibitem[\protect\citeauthoryear{{Gratton}, {Villanova}, {Lucatello},
  {Sollima}, {Geisler}, {Carretta}, {Cassisi} \& {Bragaglia}}{{Gratton}
  et~al.}{2012}]{Gratton2012}
{Gratton} R.~G.,  {Villanova} S.,  {Lucatello} S.,  {Sollima} A.,  {Geisler}
  D.,  {Carretta} E.,  {Cassisi} S.,    {Bragaglia} A.,  2012, \aap, 544, A12

\bibitem[\protect\citeauthoryear{{Grillmair}, {Freeman}, {Irwin} \&
  {Quinn}}{{Grillmair} et~al.}{1995}]{Grillmair1995}
{Grillmair} C.~J.,  {Freeman} K.~C.,  {Irwin} M.,    {Quinn} P.~J.,  1995, \aj,
  109, 2553

\bibitem[\protect\citeauthoryear{{Grillmair} \& {Johnson}}{{Grillmair} \&
  {Johnson}}{2006}]{Grillmair2006}
{Grillmair} C.~J.,  {Johnson} R.,  2006, \apjl, 639, L17

\bibitem[\protect\citeauthoryear{{Hammersley} \&
  {L{\'o}pez-Corredoira}}{{Hammersley} \&
  {L{\'o}pez-Corredoira}}{2011}]{Hammersley2011}
{Hammersley} P.~L.,  {L{\'o}pez-Corredoira} M.,  2011, \aap, 527, A6+

\bibitem[\protect\citeauthoryear{{Harris}}{{Harris}}{2010}]{Harris2010}
{Harris} W.~E.,  2010, ArXiv:1012.3224

\bibitem[\protect\citeauthoryear{{Huxor et al.}}{{Huxor et
  al.}}{2014}]{Huxor2014}
{Huxor et al.} 2014, \mnras, 442, 2165

\bibitem[\protect\citeauthoryear{{Innanen}, {Harris} \& {Webbink}}{{Innanen}
  et~al.}{1983}]{Innanen1983}
{Innanen} K.~A.,  {Harris} W.~E.,    {Webbink} R.~F.,  1983, \aj, 88, 338

\bibitem[\protect\citeauthoryear{{Juri{\'c} et al.}}{{Juri{\'c} et
  al.}}{2008}]{Juric2008}
{Juri{\'c} et al.} 2008, \apj, 673, 864

\bibitem[\protect\citeauthoryear{{Kazantzidis}, {Bullock}, {Zentner},
  {Kravtsov} \& {Moustakas}}{{Kazantzidis} et~al.}{2008}]{Kazantzidis2008}
{Kazantzidis} S.,  {Bullock} J.~S.,  {Zentner} A.~R.,  {Kravtsov} A.~V.,
  {Moustakas} L.~A.,  2008, \apj, 688, 254

\bibitem[\protect\citeauthoryear{{King}}{{King}}{1962}]{King1962}
{King} I.,  1962, \aj, 67, 471

\bibitem[\protect\citeauthoryear{{Klimentowski}, {{\L}okas}, {Kazantzidis},
  {Mayer}, {Mamon} \& {Prada}}{{Klimentowski} et~al.}{2009}]{Klimentowski2009}
{Klimentowski} J.,  {{\L}okas} E.~L.,  {Kazantzidis} S.,  {Mayer} L.,  {Mamon}
  G.~A.,    {Prada} F.,  2009, \mnras, 400, 2162

\bibitem[\protect\citeauthoryear{{Koposov et al.}}{{Koposov et
  al.}}{2012}]{Koposov2012}
{Koposov et al.} 2012, \apj, 750, 80

\bibitem[\protect\citeauthoryear{{Kunder et al.}}{{Kunder et
  al.}}{2014}]{Kunder2014}
{Kunder et al.} 2014, \aap, 572, A30

\bibitem[\protect\citeauthoryear{{K{\"u}pper}, {Kroupa}, {Baumgardt} \&
  {Heggie}}{{K{\"u}pper} et~al.}{2010}]{Kupper2010}
{K{\"u}pper} A.~H.~W.,  {Kroupa} P.,  {Baumgardt} H.,    {Heggie} D.~C.,  2010,
  \mnras, 407, 2241

\bibitem[\protect\citeauthoryear{{Kuzma}, {Da Costa} \& {Mackey}}{{Kuzma}
  et~al.}{2017}]{Kuzma2017}
{Kuzma} P.~B.,  {Da Costa} G.~S.,    {Mackey} A.~D.,  2017, ArXiv e-prints

\bibitem[\protect\citeauthoryear{{Kuzma}, {Da Costa}, {Mackey} \&
  {Roderick}}{{Kuzma} et~al.}{2016}]{Kuzma2016}
{Kuzma} P.~B.,  {Da Costa} G.~S.,  {Mackey} A.~D.,    {Roderick} T.~A.,  2016,
  \mnras, 461, 3639

\bibitem[\protect\citeauthoryear{{Law} \& {Majewski}}{{Law} \&
  {Majewski}}{2010a}]{Law2010b}
{Law} D.~R.,  {Majewski} S.~R.,  2010a, \apj, 718, 1128

\bibitem[\protect\citeauthoryear{{Law} \& {Majewski}}{{Law} \&
  {Majewski}}{2010b}]{Law2010a}
{Law} D.~R.,  {Majewski} S.~R.,  2010b, \apj, 714, 229

\bibitem[\protect\citeauthoryear{{Leaman}, {VandenBerg} \& {Mendel}}{{Leaman}
  et~al.}{2013}]{Leaman2013}
{Leaman} R.,  {VandenBerg} D.~A.,    {Mendel} J.~T.,  2013, \mnras, 436, 122

\bibitem[\protect\citeauthoryear{{Lee}, {Lee} \& {Sung}}{{Lee}
  et~al.}{2006}]{Lee2006a}
{Lee} K.~H.,  {Lee} H.~M.,    {Sung} H.,  2006, \mnras, 367, 646

\bibitem[\protect\citeauthoryear{{Lee}, {Gim} \& {Casetti-Dinescu}}{{Lee}
  et~al.}{2007}]{Lee2007}
{Lee} Y.-W.,  {Gim} H.~B.,    {Casetti-Dinescu} D.~I.,  2007, \apjl, 661, L49

\bibitem[\protect\citeauthoryear{{Leon}, {Meylan} \& {Combes}}{{Leon}
  et~al.}{2000}]{Leon2000}
{Leon} S.,  {Meylan} G.,    {Combes} F.,  2000, \aap, 359, 907

\bibitem[\protect\citeauthoryear{{Li et al.}}{{Li et al.}}{2016}]{Li2016}
{Li et al.} 2016, \apj, 817, 135

\bibitem[\protect\citeauthoryear{{L{\'o}pez-Corredoira} \&
  {Molg{\'o}}}{{L{\'o}pez-Corredoira} \&
  {Molg{\'o}}}{2014}]{Lopez-Corredoira2014}
{L{\'o}pez-Corredoira} M.,  {Molg{\'o}} J.,  2014, \aap, 567, A106

\bibitem[\protect\citeauthoryear{{Mackey et al.}}{{Mackey et
  al.}}{2010}]{Mackey2010}
{Mackey et al.} 2010, \apjl, 717, L11

\bibitem[\protect\citeauthoryear{{Mackey et al.}}{{Mackey et
  al.}}{2013}]{Mackey2013}
{Mackey et al.} 2013, \mnras, 429, 281

\bibitem[\protect\citeauthoryear{{Majewski}, {Ostheimer}, {Rocha-Pinto},
  {Patterson}, {Guhathakurta} \& {Reitzel}}{{Majewski}
  et~al.}{2004}]{Majewski2004}
{Majewski} S.~R.,  {Ostheimer} J.~C.,  {Rocha-Pinto} H.~J.,  {Patterson} R.~J.,
   {Guhathakurta} P.,    {Reitzel} D.,  2004, \apj, 615, 738

\bibitem[\protect\citeauthoryear{{Majewski}, {Skrutskie}, {Weinberg} \&
  {Ostheimer}}{{Majewski} et~al.}{2003}]{Majewski2003}
{Majewski} S.~R.,  {Skrutskie} M.~F.,  {Weinberg} M.~D.,    {Ostheimer} J.~C.,
  2003, \apj, 599, 1082

\bibitem[\protect\citeauthoryear{{Mar{\'{\i}}n-Franch et
  al.}}{{Mar{\'{\i}}n-Franch et al.}}{2009}]{Marin-Franch2009}
{Mar{\'{\i}}n-Franch et al.} 2009, \apj, 694, 1498

\bibitem[\protect\citeauthoryear{{Marino et al.}}{{Marino et
  al.}}{2014}]{Marino2014}
{Marino et al.} 2014, \mnras, 442, 3044

\bibitem[\protect\citeauthoryear{{Martin}, {Ibata}, {Bellazzini}, {Irwin},
  {Lewis} \& {Dehnen}}{{Martin} et~al.}{2004}]{Martin2004}
{Martin} N.~F.,  {Ibata} R.~A.,  {Bellazzini} M.,  {Irwin} M.~J.,  {Lewis}
  G.~F.,    {Dehnen} W.,  2004, \mnras, 348, 12

\bibitem[\protect\citeauthoryear{{Martin}, {Ibata} \& {Irwin}}{{Martin}
  et~al.}{2007}]{Martin2007}
{Martin} N.~F.,  {Ibata} R.~A.,    {Irwin} M.,  2007, \apjl, 668, L123

\bibitem[\protect\citeauthoryear{{Mart{\'{\i}}nez-Delgado}, {Butler}, {Rix},
  {Franco}, {Pe{\~n}arrubia}, {Alfaro} \& {Dinescu}}{{Mart{\'{\i}}nez-Delgado}
  et~al.}{2005}]{Martinez-Delgado2005}
{Mart{\'{\i}}nez-Delgado} D.,  {Butler} D.~J.,  {Rix} H.-W.,  {Franco} V.~I.,
  {Pe{\~n}arrubia} J.,  {Alfaro} E.~J.,    {Dinescu} D.~I.,  2005, \apj, 633,
  205

\bibitem[\protect\citeauthoryear{{Mart{\'{\i}}nez-Delgado}, {Zinn}, {Carrera}
  \& {Gallart}}{{Mart{\'{\i}}nez-Delgado} et~al.}{2002}]{Martinez-Delgado2002}
{Mart{\'{\i}}nez-Delgado} D.,  {Zinn} R.,  {Carrera} R.,    {Gallart} C.,
  2002, \apjl, 573, L19

\bibitem[\protect\citeauthoryear{{Mateu}, {Vivas}, {Zinn}, {Miller} \&
  {Abad}}{{Mateu} et~al.}{2009}]{Mateu2009}
{Mateu} C.,  {Vivas} A.~K.,  {Zinn} R.,  {Miller} L.~R.,    {Abad} C.,  2009,
  \aj, 137, 4412

\bibitem[\protect\citeauthoryear{{Milone}, {Marino}, {Piotto}, {Renzini},
  {Bedin}, {Anderson}, {Cassisi}, {D'Antona}, {Bellini}, {Jerjen},
  {Pietrinferni} \& {Ventura}}{{Milone} et~al.}{2015}]{Milone2015}
{Milone} A.~P.,  {Marino} A.~F.,  {Piotto} G.,  {Renzini} A.,  {Bedin} L.~R.,
  {Anderson} J.,  {Cassisi} S.,  {D'Antona} F.,  {Bellini} A.,  {Jerjen} H.,
  {Pietrinferni} A.,    {Ventura} P.,  2015, \apj, 808, 51

\bibitem[\protect\citeauthoryear{{Milone}, {Stetson}, {Piotto}, {Bedin},
  {Anderson}, {Cassisi} \& {Salaris}}{{Milone} et~al.}{2009}]{Milone2009}
{Milone} A.~P.,  {Stetson} P.~B.,  {Piotto} G.,  {Bedin} L.~R.,  {Anderson} J.,
   {Cassisi} S.,    {Salaris} M.,  2009, \aap, 503, 755

\bibitem[\protect\citeauthoryear{{Milone et al.}}{{Milone et
  al.}}{2008}]{Milone2008}
{Milone et al.} 2008, \apj, 673, 241

\bibitem[\protect\citeauthoryear{{Miyamoto} \& {Nagai}}{{Miyamoto} \&
  {Nagai}}{1975}]{Miyamoto1975}
{Miyamoto} M.,  {Nagai} R.,  1975, \pasj, 27, 533

\bibitem[\protect\citeauthoryear{{Moitinho}, {V{\'a}zquez}, {Carraro}, {Baume},
  {Giorgi} \& {Lyra}}{{Moitinho} et~al.}{2006}]{Moitinho2006}
{Moitinho} A.,  {V{\'a}zquez} R.~A.,  {Carraro} G.,  {Baume} G.,  {Giorgi}
  E.~E.,    {Lyra} W.,  2006, \mnras, 368, L77

\bibitem[\protect\citeauthoryear{{Momany}, {Zaggia}, {Gilmore}, {Piotto},
  {Carraro}, {Bedin} \& {de Angeli}}{{Momany} et~al.}{2006}]{Momany2006}
{Momany} Y.,  {Zaggia} S.,  {Gilmore} G.,  {Piotto} G.,  {Carraro} G.,  {Bedin}
  L.~R.,    {de Angeli} F.,  2006, \aap, 451, 515

\bibitem[\protect\citeauthoryear{{Montuori}, {Capuzzo-Dolcetta}, {Di Matteo},
  {Lepinette} \& {Miocchi}}{{Montuori} et~al.}{2007}]{Montuori2007}
{Montuori} M.,  {Capuzzo-Dolcetta} R.,  {Di Matteo} P.,  {Lepinette} A.,
  {Miocchi} P.,  2007, \apj, 659, 1212

\bibitem[\protect\citeauthoryear{{Morganson et al.}}{{Morganson et
  al.}}{2016}]{Morganson2016}
{Morganson et al.} 2016, \apj, 825, 140

\bibitem[\protect\citeauthoryear{{Myeong}, {Jerjen}, {Mackey} \& {Da
  Costa}}{{Myeong} et~al.}{2017}]{Myeong2017}
{Myeong} G.~C.,  {Jerjen} H.,  {Mackey} D.,    {Da Costa} G.~S.,  2017, ArXiv
  e-prints

\bibitem[\protect\citeauthoryear{{Natarajan} \& {Sikivie}}{{Natarajan} \&
  {Sikivie}}{2007}]{Natarajan2007}
{Natarajan} A.,  {Sikivie} P.,  2007, \prd, 76, 023505

\bibitem[\protect\citeauthoryear{{Navarrete}, {Belokurov} \&
  {Koposov}}{{Navarrete} et~al.}{2017}]{Navarrete2017}
{Navarrete} C.,  {Belokurov} V.,    {Koposov} S.~E.,  2017, \apjl, 841, L23

\bibitem[\protect\citeauthoryear{{Navin et al.}}{{Navin et
  al.}}{2015}]{Navin2015}
{Navin et al.} 2015, \mnras, 453, 531

\bibitem[\protect\citeauthoryear{{Newberg et al.}}{{Newberg et
  al.}}{2002}]{Newberg2002}
{Newberg et al.} 2002, \apj, 569, 245

\bibitem[\protect\citeauthoryear{{Odenkirchen}, {Grebel}, {Dehnen}, {Rix},
  {Yanny}, {Newberg}, {Rockosi}, {Mart{\'{\i}}nez-Delgado}, {Brinkmann} \&
  {Pier}}{{Odenkirchen} et~al.}{2003}]{Odenkirchen2003}
{Odenkirchen} M.,  {Grebel} E.~K.,  {Dehnen} W.,  {Rix} H.,  {Yanny} B.,
  {Newberg} H.~J.,  {Rockosi} C.~M.,  {Mart{\'{\i}}nez-Delgado} D.,
  {Brinkmann} J.,    {Pier} J.~R.,  2003, \aj, 126, 2385

\bibitem[\protect\citeauthoryear{{Odenkirchen et al.}}{{Odenkirchen et
  al.}}{2001}]{Odenkirchen2001}
{Odenkirchen et al.} 2001, \apjl, 548, L165

\bibitem[\protect\citeauthoryear{{Olszewski}, {Saha}, {Knezek}, {Subramaniam},
  {de Boer} \& {Seitzer}}{{Olszewski} et~al.}{2009}]{Olszewski2009}
{Olszewski} E.~W.,  {Saha} A.,  {Knezek} P.,  {Subramaniam} A.,  {de Boer} T.,
    {Seitzer} P.,  2009, \aj, 138, 1570

\bibitem[\protect\citeauthoryear{{Pawlowski}, {Pflamm-Altenburg} \&
  {Kroupa}}{{Pawlowski} et~al.}{2012}]{Pawlowski2012}
{Pawlowski} M.~S.,  {Pflamm-Altenburg} J.,    {Kroupa} P.,  2012, \mnras, 423,
  1109

\bibitem[\protect\citeauthoryear{{Pe{\~n}arrubia}, {Mart{\'{\i}}nez-Delgado},
  {Rix}, {G{\'o}mez-Flechoso}, {Munn}, {Newberg}, {Bell}, {Yanny}, {Zucker} \&
  {Grebel}}{{Pe{\~n}arrubia} et~al.}{2005}]{Penarrubia2005}
{Pe{\~n}arrubia} J.,  {Mart{\'{\i}}nez-Delgado} D.,  {Rix} H.~W.,
  {G{\'o}mez-Flechoso} M.~A.,  {Munn} J.,  {Newberg} H.,  {Bell} E.~F.,
  {Yanny} B.,  {Zucker} D.,    {Grebel} E.~K.,  2005, \apj, 626, 128

\bibitem[\protect\citeauthoryear{{Piotto}, {Bedin}, {Anderson}, {King},
  {Cassisi}, {Milone}, {Villanova}, {Pietrinferni} \& {Renzini}}{{Piotto}
  et~al.}{2007}]{Piotto2007}
{Piotto} G.,  {Bedin} L.~R.,  {Anderson} J.,  {King} I.~R.,  {Cassisi} S.,
  {Milone} A.~P.,  {Villanova} S.,  {Pietrinferni} A.,    {Renzini} A.,  2007,
  \apjl, 661, L53

\bibitem[\protect\citeauthoryear{{Piotto}, {Milone}, {Anderson}, {Bedin},
  {Bellini}, {Cassisi}, {Marino}, {Aparicio} \& {Nascimbeni}}{{Piotto}
  et~al.}{2012}]{Piotto2012}
{Piotto} G.,  {Milone} A.~P.,  {Anderson} J.,  {Bedin} L.~R.,  {Bellini} A.,
  {Cassisi} S.,  {Marino} A.~F.,  {Aparicio} A.,    {Nascimbeni} V.,  2012,
  \apj, 760, 39

\bibitem[\protect\citeauthoryear{{Piotto et al.}}{{Piotto et
  al.}}{2015}]{Piotto2015}
{Piotto et al.} 2015, \aj, 149, 91

\bibitem[\protect\citeauthoryear{{Price-Whelan}, {Johnston}, {Sheffield},
  {Laporte} \& {Sesar}}{{Price-Whelan} et~al.}{2015}]{Price-Whelan2015}
{Price-Whelan} A.~M.,  {Johnston} K.~V.,  {Sheffield} A.~A.,  {Laporte}
  C.~F.~P.,    {Sesar} B.,  2015, \mnras, 452, 676

\bibitem[\protect\citeauthoryear{{Purcell}, {Bullock}, {Tollerud}, {Rocha} \&
  {Chakrabarti}}{{Purcell} et~al.}{2011}]{Purcell2011}
{Purcell} C.~W.,  {Bullock} J.~S.,  {Tollerud} E.~J.,  {Rocha} M.,
  {Chakrabarti} S.,  2011, \nat, 477, 301

\bibitem[\protect\citeauthoryear{{Robin}, {Reyl{\'e}}, {Derri{\`e}re} \&
  {Picaud}}{{Robin} et~al.}{2003}]{Robin2003}
{Robin} A.~C.,  {Reyl{\'e}} C.,  {Derri{\`e}re} S.,    {Picaud} S.,  2003,
  \aap, 409, 523

\bibitem[\protect\citeauthoryear{{Rocha-Pinto}, {Majewski}, {Skrutskie},
  {Crane} \& {Patterson}}{{Rocha-Pinto} et~al.}{2004}]{Rocha-Pinto2004}
{Rocha-Pinto} H.~J.,  {Majewski} S.~R.,  {Skrutskie} M.~F.,  {Crane} J.~D.,
  {Patterson} R.~J.,  2004, \apj, 615, 732

\bibitem[\protect\citeauthoryear{{Rockosi et al.}}{{Rockosi et
  al.}}{2002}]{Rockosi2002}
{Rockosi et al.} 2002, \aj, 124, 349

\bibitem[\protect\citeauthoryear{{Sbordone et al.}}{{Sbordone et
  al.}}{2015}]{Sbordone2015}
{Sbordone et al.} 2015, \aap, 579, A104

\bibitem[\protect\citeauthoryear{{Schlafly} \& {Finkbeiner}}{{Schlafly} \&
  {Finkbeiner}}{2011}]{Schlafly2011}
{Schlafly} E.~F.,  {Finkbeiner} D.~P.,  2011, \apj, 737, 103

\bibitem[\protect\citeauthoryear{{Sch{\"o}nrich}, {Binney} \&
  {Dehnen}}{{Sch{\"o}nrich} et~al.}{2010}]{Schonrich2010}
{Sch{\"o}nrich} R.,  {Binney} J.,    {Dehnen} W.,  2010, \mnras, 403, 1829

\bibitem[\protect\citeauthoryear{{Sheffield}, {Johnston}, {Majewski}, {Damke},
  {Richardson}, {Beaton} \& {Rocha-Pinto}}{{Sheffield}
  et~al.}{2014}]{Sheffield2014}
{Sheffield} A.~A.,  {Johnston} K.~V.,  {Majewski} S.~R.,  {Damke} G.,
  {Richardson} W.,  {Beaton} R.,    {Rocha-Pinto} H.~J.,  2014, \apj, 793, 62

\bibitem[\protect\citeauthoryear{{Simion}, {Belokurov}, {Irwin} \&
  {Koposov}}{{Simion} et~al.}{2014}]{Simion2014}
{Simion} I.~T.,  {Belokurov} V.,  {Irwin} M.,    {Koposov} S.~E.,  2014,
  \mnras, 440, 161

\bibitem[\protect\citeauthoryear{{Slater et al.}}{{Slater et
  al.}}{2014}]{Slater2014}
{Slater et al.} 2014, \apj, 791, 9

\bibitem[\protect\citeauthoryear{{Sollima}, {Gratton}, {Carballo-Bello},
  {Mart{\'{\i}}nez-Delgado}, {Carretta}, {Bragaglia}, {Lucatello} \&
  {Pe{\~n}arrubia}}{{Sollima} et~al.}{2012}]{Sollima2012}
{Sollima} A.,  {Gratton} R.~G.,  {Carballo-Bello} J.~A.,
  {Mart{\'{\i}}nez-Delgado} D.,  {Carretta} E.,  {Bragaglia} A.,  {Lucatello}
  S.,    {Pe{\~n}arrubia} J.,  2012, \mnras, 426, 1137

\bibitem[\protect\citeauthoryear{{Sollima}, {Valls-Gabaud}, {Martinez-Delgado},
  {Fliri}, {Pe{\~n}arrubia} \& {Hoekstra}}{{Sollima}
  et~al.}{2011}]{Sollima2011}
{Sollima} A.,  {Valls-Gabaud} D.,  {Martinez-Delgado} D.,  {Fliri} J.,
  {Pe{\~n}arrubia} J.,    {Hoekstra} H.,  2011, \apjl, 730, L6

\bibitem[\protect\citeauthoryear{{Steinmetz}}{{Steinmetz}}{2006}]{Steinmetz2006}
{Steinmetz} 2006, \aj, 132, 1645

\bibitem[\protect\citeauthoryear{{Stetson}}{{Stetson}}{1987}]{Stetson1987}
{Stetson} P.~B.,  1987, \pasp, 99, 191

\bibitem[\protect\citeauthoryear{{Trager}, {King} \& {Djorgovski}}{{Trager}
  et~al.}{1995}]{Trager1995}
{Trager} S.~C.,  {King} I.~R.,    {Djorgovski} S.,  1995, \aj, 109, 218

\bibitem[\protect\citeauthoryear{{Valdes}, {Gruendl} \& {DES Project}}{{Valdes}
  et~al.}{2014}]{Valdes2014}
{Valdes} F.,  {Gruendl} R.,    {DES Project} 2014, in {Manset} N.,  {Forshay}
  P.,  eds, Astronomical Data Analysis Software and Systems XXIII Vol.~485 of
  Astronomical Society of the Pacific Conference Series, {The DECam Community
  Pipeline}.
p.~379

\bibitem[\protect\citeauthoryear{{Veljanoski et al.}}{{Veljanoski et
  al.}}{2014}]{Veljanoski2014}
{Veljanoski et al.} 2014, \mnras, 442, 2929

\bibitem[\protect\citeauthoryear{{Vivas}, {Zinn}, {Farmer}, {Duffau} \&
  {Ping}}{{Vivas} et~al.}{2016}]{Vivas2016}
{Vivas} A.~K.,  {Zinn} R.,  {Farmer} J.,  {Duffau} S.,    {Ping} Y.,  2016,
  \apj, 831, 165

\bibitem[\protect\citeauthoryear{{Wagner-Kaiser}, {Sarajedini}, {von Hippel},
  {Stenning}, {van Dyk}, {Jeffery}, {Robinson}, {Stein}, {Anderson} \&
  {Jefferys}}{{Wagner-Kaiser} et~al.}{2017}]{Wagner-Kaiser2017}
{Wagner-Kaiser} R.,  {Sarajedini} A.,  {von Hippel} T.,  {Stenning} D.~C.,
  {van Dyk} D.~A.,  {Jeffery} E.,  {Robinson} E.,  {Stein} N.,  {Anderson} J.,
    {Jefferys} W.~H.,  2017, \mnras, 468, 1038

\bibitem[\protect\citeauthoryear{{Wilson}}{{Wilson}}{1975}]{Wilson1975}
{Wilson} C.~P.,  1975, \aj, 80, 175

\bibitem[\protect\citeauthoryear{{Witham}, {Knigge}, {Drew}, {Greimel},
  {Steeghs}, {G{\"a}nsicke}, {Groot} \& {Mampaso}}{{Witham}
  et~al.}{2008}]{Witham2008}
{Witham} A.~R.,  {Knigge} C.,  {Drew} J.~E.,  {Greimel} R.,  {Steeghs} D.,
  {G{\"a}nsicke} B.~T.,  {Groot} P.~J.,    {Mampaso} A.,  2008, \mnras, 384,
  1277

\bibitem[\protect\citeauthoryear{{Xu}, {Newberg}, {Carlin}, {Liu}, {Deng},
  {Li}, {Sch{\"o}nrich} \& {Yanny}}{{Xu} et~al.}{2015}]{Xu2015}
{Xu} Y.,  {Newberg} H.~J.,  {Carlin} J.~L.,  {Liu} C.,  {Deng} L.,  {Li} J.,
  {Sch{\"o}nrich} R.,    {Yanny} B.,  2015, \apj, 801, 105

\bibitem[\protect\citeauthoryear{{Yanny}, {Newberg}, {Grebel}, {Kent},
  {Odenkirchen}, {Rockosi}, {Schlegel}, {Subbarao}, {Brinkmann}, {Fukugita},
  {Ivezic}, {Lamb}, {Schneider} \& {York}}{{Yanny} et~al.}{2003}]{Yanny2003}
{Yanny} B.,  {Newberg} H.~J.,  {Grebel} E.~K.,  {Kent} S.,  {Odenkirchen} M.,
  {Rockosi} C.~M.,  {Schlegel} D.,  {Subbarao} M.,  {Brinkmann} J.,  {Fukugita}
  M.,  {Ivezic} {\v Z}.,  {Lamb} D.~Q.,  {Schneider} D.~P.,    {York} D.~G.,
  2003, \apj, 588, 824

\bibitem[\protect\citeauthoryear{{Zaritsky}, {Crnojevi{\'c}} \&
  {Sand}}{{Zaritsky} et~al.}{2016}]{Zaritsky2016}
{Zaritsky} D.,  {Crnojevi{\'c}} D.,    {Sand} D.~J.,  2016, \apjl, 826, L9

\bibitem[\protect\citeauthoryear{{Zinn}}{{Zinn}}{1993}]{Zinn1993}
{Zinn} R.,  1993, in {G.~H.~Smith \& J.~P.~Brodie} ed., The Globular
  Cluster-Galaxy Connection Vol.~48 of Astronomical Society of the Pacific
  Conference Series, {The Galactic Halo Cluster Systems: Evidence for
  Accretion}.
p.~38

\bibitem[\protect\citeauthoryear{{Zoccali}, {Pancino}, {Catelan}, {Hempel},
  {Rejkuba} \& {Carrera}}{{Zoccali} et~al.}{2009}]{Zoccali2009}
{Zoccali} M.,  {Pancino} E.,  {Catelan} M.,  {Hempel} M.,  {Rejkuba} M.,
  {Carrera} R.,  2009, \apjl, 697, L22

\end{thebibliography}

\bsp	
\label{lastpage}
\end{document}